\documentclass[3p, 10pt, twocolumn]{elsarticle}
\usepackage{multicol}
\usepackage{color}
\usepackage{comment}
\usepackage{float}
\usepackage{xfrac}
\usepackage{soul}

\usepackage{amssymb, amsmath}
\usepackage[nodots]{numcompress}

\usepackage{placeins}

\journal{Journal of Quantitative Spectroscopy and Radiative Transfer}

\allowdisplaybreaks

\begin{document}

\setstcolor{red}

\begin{frontmatter}

%% Title, authors and addresses

%% use the tnoteref command within \title for footnotes;
%% use the tnotetext command for the associated footnote;
%% use the fnref command within \author or \address for footnotes;
%% use the fntext command for the associated footnote;
%% use the corref command within \author for corresponding author footnotes;
%% use the cortext command for the associated footnote;
%% use the ead command for the email address,
%% and the form \ead[url] for the home page:
%%
%% \title{Title\tnoteref{label1}}
%% \tnotetext[label1]{}
%% \author{Name\corref{cor1}\fnref{label2}}
%% \ead{email address}
%% \ead[url]{home page}
%% \fntext[label2]{}
%% \cortext[cor1]{}
%% \address{Address\fnref{label3}}
%% \fntext[label3]{}

\title{The equilibrium-diffusion limit for radiation hydrodynamics}

%% use optional labels to link authors explicitly to addresses:
%% \author[label1,label2]{<author name>}
%% \address[label1]{<address>}
%% \address[label2]{<address>}

\author[lanl_xdiv]{J.M.~Ferguson}

\author[tamu_nuen]{J.E.~Morel}

\author[lanl_ccs2]{R.B.~Lowrie}

\address[lanl_xdiv]{XCP-Division, Los Alamos National Laboratory, Los Alamos, NM 87545, USA}
\address[tamu_nuen]{Department of Nuclear Engineering, Texas A \& M University, College Station, TX 77843, USA}
\address[lanl_ccs2]{CCS-Division, Los Alamos National Laboratory, Los Alamos, NM 87545, USA}

\begin{comment}
  (While this comment was made at the beginning of Section 4 I moved it here because it needs to be applied to various parts of the paper.)
  JEM: We never discuss the fact that our results only apply on the asymptotic "interior" of the domain, i.e., far from discontinuities or boundary layers. 
\end{comment}

\begin{abstract}
%% Text of abstract
The equilibrium-diffusion approximation (EDA) is used to describe certain radiation-hydrodynamic (RH) environments.
When this is done the RH equations reduce to a simplified set of equations.
The EDA can be derived by asymptotically analyzing the full set of RH equations in the equilibrium-diffusion limit.
We derive the EDA this way and show that it and the associated set of simplified equations are both first-order accurate with transport corrections occurring at second order.
Having established the EDA's first-order accuracy we then analyze the grey nonequilibrium-diffusion approximation and the grey Eddington approximation and show that they both preserve this first-order accuracy.
Further, these approximations preserve the EDA's first-order accuracy when made in either the comoving-frame (CMF) or the lab-frame (LF).
\begin{comment}
  JEM: I think it would aid clarity to say that the sources are identical to (beta^2$) in the two frames when expressed in terms of the energy density, flux and pressure, but that the CMF and LF pressures themselves are not equivalent.
  JMF: By explicitly mentioning ``the energy density, flux and pressure'' are you trying to isolate the comment to the derivatives only?  The equivalence applies to the right-hand side of the equations as well.
\end{comment}
% During our analysis we found that the radiation-source equations under the Eddington approximation are equivalent in the two frames when neglecting ${\cal O}(\beta^2)$ terms; of course, the radiation pressures are not strictly equivalent.
While analyzing the Eddington approximation, we found that the CMF and LF radiation-source equations are equivalent when neglecting ${\cal O}(\beta^2)$ terms and compared in the LF.
Of course, the radiation pressures are not equivalent.
It is expected that simplified physical models and numerical discretizations of the RH equations that do not preserve this first-order accuracy will not retain the correct equilibrium-diffusion solutions.
As a practical example, we show that nonequilibrium-diffusion radiative-shock solutions devolve to equilibrium-diffusion solutions when the asymptotic parameter is small.
\end{abstract}

\begin{keyword}
%% keywords here, in the form: keyword \sep keyword
asymptotics \sep equilibrium diffusion \sep radiation transport \sep radiation hydrodynamics \sep grey nonequilibrium-diffusion approximation \sep grey Eddington approximation, radiative-shock solutions
%% MSC codes here, in the form: \MSC code \sep code
%% or \MSC[2008] code \sep code (2000 is the default)

\end{keyword}

\end{frontmatter}

\section{Introduction}
\label{sec:intro}
Radiation hydrodynamics (RH) describes how interactions between radiation and matter affect the thermodynamic states, and potentially, the dynamic flow characteristics of the matter-radiation system.
Unfortunately, the full set of RH equations are computationally expensive and numerically difficult to solve, and various model approximations have been developed to aid their solution \cite{ZR2002, MM1999, Castor2007}.
The Euler equations are typically assumed to provide a sufficient model of the material's hydrodynamic response since the photon mean-free-path is generally much longer than the mean-free-path between material interactions, so that material viscosity and heat conduction may be neglected.
\begin{comment}
  JEM: The flux is not time time independent because the gradient of the energy density is not time independent.  The time derivative of the flux is set to zero in the momentum equation because it is smaller than the terms that are retained, but it is not zero.
  JMF: Fixed.
\end{comment}
For the radiation it is common to make the following assumptions, either independently or together: the system is absorption dominated, the system's size is large compared to the photon mean-free-path, the radiation is in thermal equilibrium with the material, the radiation flux is diffusive,
% and time independent,
and the radiation pressure is isotropic.
Taken together these assumptions are called the equilibrium diffusion approximation (EDA) \cite{MM1999}.
When the EDA applies to a physical system it can be modeled by a simpler set of equations than the full set of RH equations.
This simplified set of RH equations provides a reasonable description of stellar structure \cite{Chandrasekhar1960}, high-temperature environments \cite{ZR2002},  fusion dominated energy sources \cite{GL1975, MtV2002}, a variety of astrophysical settings \cite{MM1999, Shu1991radiation, Shu1991gas}, and high-energy-density-physics \cite{Drake2007}.

As shown by Lowrie, Morel and Hittinger \cite{LMH1999}, the EDA can be derived via an asymptotic expansion of the RH equations, which is described as follows.
\begin{comment}
  JEM: The step of non-dimensionalizing the equations is missing.
  JMF: Fixed.
\end{comment}
% First, the RH solution is expressed as an infinite power-series expansion in a small parameter, $\epsilon$.
First, an ansatz for the RH solution is made and the solution is expressed as an infinite power-series expansion in a small parameter, $\epsilon$.
Then the RH equations are nondimensionalized.
Next, nondimensional parameters that arise in the RH equations are chosen to scale by particular powers of $\epsilon$, which reflects the physical importance of those parameters, and effectively defines physical and mathematical limits.
Finally, an infinite hierarchy of equations is obtained by equating all terms multiplied by the same power of $\epsilon$.
\begin{comment}
  JEM: This phrase seems to be saying that the scalings are chosen to drop terms of order epsilon on higher.  No terms are actually dropped.  I would eliminate the phrase.
  JMF: Fixed.
\end{comment}
This hierarchy is used to define the equilibrium-diffusion limit (EDL) since the scalings are chosen to ensure that the zeroth order RH solution satisfies the EDA.
% by dropping terms that are of order $\epsilon$, or smaller.
Thus, in the limit as $\epsilon$ approaches zero the RH solution transitions to the EDA solution.
In this sense, the EDA is said to be accurate or exact to zeroth order in the EDL.
However, it is possible for an approximation to be exact to higher than zeroth order.
This has been shown to be true for the EDA by Lowrie, Morel and Hittinger \cite{LMH1999}, but they only analyzed the simplified RH equations at zeroth order and they did not determine at what order transport corrections modified the EDA.
The main contribution of this paper is to show that the EDA and the simplified RH equations are both exact to first order in the EDL, and that transport corrections occur at second order.
It is important to be clear that the EDA is a physical approximation to the full set of RH equations, whereas the asymptotic expansion of the RH equations is exact at each asymptotic level.
If we summed over all asymptotic orders then the original transport content would be regained with no loss of information.

Asymptotic expansions have two very practical applications that are not necessarily obvious.
The first is to demonstrate that analytic approximations to the full RH equations give proper accuracy in the EDL.
For instance, one would expect the grey nonequilibrium-diffusion approximation of the RH equations to be “properly” accurate in the EDL.
To determine if this is so, one performs an asymptotic expansion of these approximate RH equations.
% in the EDL.
If this expansion yields the EDA to first order, then the solution of the approximate RH equations will properly transition to the EDA solution as $\epsilon \rightarrow 0$.
Otherwise, the solution will approach the EDA solution at the wrong rate, or in the worst case, not approach it at all.
The second application relates to numerical discretizations of the RH equations and their approximations.
Numerical schemes are said to be “asymptotic preserving” when they give proper accuracy in an asymptotic limit at an appropriate computational cost.
Numerical discretizations that are consistent, i.e., converge to the proper analytic solution as the mesh is refined, are not necessarily asymptotic preserving.
To determine if a numerical discretization is asymptotic preserving, one performs the same asymptotic expansion for the discrete equations that was performed for the analytic equations.
The discrete asymptotic expansion must approximate the analytic expansion in certain specific ways to be asymptotic preserving.
If a consistent scheme is not asymptotic preserving, one can obtain accurate solutions in problems closely approaching the asymptotic limit, but the computational cost will generally be prohibitive.
In this paper we analyze two approximations of the RH equations, but we do not analyze any discretization schemes.
For future work on numerical discretizations, this paper defines the sense in which discretization schemes should, ideally, preserve the EDA.
Specifically, to be asymptotic preserving, a numerical discretization must produce the EDA and a discretized set of its simplified RH equations through first order in the EDL.

Similar analyses have been performed for neutron transport \cite{HM1975, L1975} and radiative transfer \cite{LPB1983, Morel2000}, which have shown that their analytic diffusion approximations are first-order accurate.
Simplified physical models and numerical discretizations of these theories which fail to preserve the diffusion limit typically fail to obtain accurate diffusion solutions \cite{MalvagiePomraning1991, LK1974, L1975, HM1975, LMM1987, LM1989, AdamsWareingWalters1998, Adams2001, LPB1983, Morel2000, LP1980, MorelWareingSmith1996, AN1998, OAH2000, DL2004, D2011}.
\begin{comment}
  JEM: I find this phrase (physical-model simplifications) awkward.
       Simplified physical models seems more natural to me?
  JMF: Fixed.
\end{comment}
\begin{comment}
  JEM: This phrase only applies to numerical discretizations - not physical model simplifications.
       It could be replaced by "or else they will fail in some sense in the EDL."
  JMF: Fixed.  Rob had previously recommended a fix for the last two sentences of Section 3, and that fix applies here as well.
\end{comment}
Therefore, we expect that simplified physical models and numerical discretizations of the RH equations should preserve this asymptotic behavior in order to reasonably obtain correct equilibrium-diffusion solutions.
When a physical model or numerical discretization preserves some diffusion limit they are generally referred to as being asymptotic preserving \cite{J1999}.
% Models that do not have the correct asymptotic behavior may not attain the correct solution in or near the EDL, while improper discretizations may require an unreasonable computational cost.
% We similarly expect simplified models and numerical discretizations of the RH equations should also preserve the EDA's first-order accuracy, or else we expect that they will fail to obtain accurate equilibrium-diffusion solutions at a reasonable computational cost.
Previous analyses \cite{LMM1987, LM1989, LPB1983, MalvagiePomraning1991} for neutron transport and radiative transfer have discussed the effects of initial and boundary conditions, as well as boundary layers, on the asymptotic results.
However, in this paper we restrict our analysis to the interior solution sufficiently late in time and far from any boundaries so that their effects on the analysis may be neglected.
An analysis including the initial and boundary conditions, and potentially boundary layers, should be the subject of future work.

The rest of this paper is organized as follows:
In Section \ref{sec:equations}, the lab-frame (LF) RH equations and the EDA are presented.
In Section \ref{sec:asymptotic_analysis}, the main result of this paper is presented: the EDA and the simplified RH equations are first-order accurate in the EDL with transport corrections occurring at second-order.
The derivation of these results is given in \ref{app:app3A}.
The EDA is a physical limit of the RH equations and we claim that its first-order accuracy should be preserved by simplified models of the RH equations.
In Section \ref{sec:LF_CMF}, the grey nonequilibrium-diffusion approximation is analyzed and shown to preserve the EDA's first-order accuracy.
Further, we show that the nonequilibrium-diffusion approximation may be applied in either the comoving-frame (CMF) or the LF and the EDA's first-order accuracy is preserved in both cases.
As a practical example we show that a particular nonequilibrium-diffusion radiative-shock solution \cite{LE2008} devolves to the appropriate equilibrium-diffusion solution \cite{LR2007} when the asymptotic parameter is small.
Then, the grey Eddington approximation is analyzed and shown to also preserve the EDA's first-order accuracy.
Again, the Eddington approximation may be applied in the CMF or the LF and the EDA's first-order accuracy is preserved in both cases.
\begin{comment}
  JEM: I think this is misleading for reasons previously discussed. (see comments in the abstract)
  JFM: Restructured sentence and added a comment that the pressures are not equivalent.
\end{comment}
% Further, we show that the radiation-source equations under the Eddington approximation are equivalent in the two frames when neglecting ${\cal O}(\beta^2)$ terms.
Further, we show that when applying the Eddington approximation in the CMF or the LF the radiation-source equations are equivalent when neglecting ${\cal O}(\beta^2)$ terms.
Of course, the radiation pressures are not equivalent, but they only differ by a symmetric-traceless term of ${\cal O}(\beta)$.
The reason for analyzing these two approximations is because the nonequilibrium-diffusion approximation only applies to optically-thick systems, and is only slightly less restrictive than the EDA.
In contradistinction, the Eddington approximation may be applied to highly rarefied systems where diffusion does not apply and which may be out of thermal equilibrium.
Section \ref{sec:summary} closes the main body of this paper with a summary and recommendations for future work.
In \ref{app:app2C} we derive the ${\cal O}(\beta^2)$ LF radiation-transport (RT) equation including scattering terms.
We believe this is the first time that this equation has been presented in the literature when retaining terms through ${\cal O}(\beta^2)$.
In \ref{app:app3A} we use results from \ref{app:app2C} to derive the radiation intensity through ${\cal O}(\epsilon^2)$.
This is used to derive the zeroth- and first-order radiation variables and the second-order radiation flux, and to show that the radiation energy and pressure contain transport corrections at second order.
\section{The RH equations and the EDA}
\label{sec:equations}
We present the LF RH equations in Subsection \ref{subsec:LF_RH_equations},
% and derive
as well as the radiation-energy and radiation-momentum sources
% as well as
and the frequency-integrated angular moments of the radiation intensity.
In Subsection \ref{subsec:CMF_EDA}, we review the assumptions on which the EDA relies and then we present the LF EDA and its simplified RH equations.
\subsection{The RH equations and radiation variables}
\label{subsec:LF_RH_equations}
The RH equations used here are the Euler equations coupled to the radiation momentum and energy sources, and the angle- and frequency-dependent RT equation:
\begin{subequations}
\label{eqs:RH_equations}
% \begin{linenomath}
  \begin{gather}
    \partial_t \rho + \partial_i \left( \rho \, u_i \right) = 0 \, , \label{eq:mass_conservation} \\
    \partial_t \left( \rho \, u_i \right) + \partial_j \left( \rho \, u_i u_j + p_{ij} \right) = - S_{\text{rp}, i} \, , \label{eq:momentum_conservation} \\
    \partial_t E + \partial_i \left[ u_j \left( E_{ij} + p_{ij} \right) \right] = - S_{\text{re}} \, , \label{eq:energy_conservation} \\
    \frac{1}{c} \, \partial_t I_{\nu} + \Omega_i \, \partial_i I_{\nu} = Q_{\nu} \, , \label{eq:LF_RT}
  \end{gather}
% \end{linenomath}
where
% \begin{linenomath}
  \begin{multline}
  \label{eq:LF_RT_source}
    Q_{\nu} = - \left( \frac{\nu_{\text{o}}}{\nu} \right) \sigma_{\text{t}, \nu_{\text{o}}} \, I_{\nu} + \left( \frac{\nu}{\nu_{\text{o}}} \right)^2 \sigma_{\text{a}, \nu_{\text{o}}} \, B_{\nu_{\text{o}}}(T) \\
    + \left( \frac{\nu}{\nu_{\text{o}}} \right)^2 \frac{\sigma_{\text{s}}}{4 \pi} \int_{4 \pi} \left( \frac{\nu_{\text{o}}}{\nu^{\, \prime}} \right) I_{\nu^{\, \prime}} \left( \Omega^{\, \prime} \right) d\Omega^{\, \prime} \, ,
  \end{multline}
% \end{linenomath}
\end{subequations}
is the angle- and frequency-dependent LF radiation source.
The angle-dependence resides in the frequency ratios which are relativistically exact through all orders in $\beta \equiv u /c$, as well as the angle- and frequency-dependent LF radiation intensity, $I_{\nu} = I_{\nu}(\Omega)$, and $\Omega_i$ is the LF photon direction of flight.
The time and space derivatives, $\partial_t \equiv \partial / \partial t$ and $\partial_i \equiv \partial x^i$, are with respect to LF time and space, where $x^i \in \{x, y, z \}$, and we are using the Einstein summation convention.
Further, $\rho$ is the mass density, $u_i$ is the material velocity, $p_{ij}$ is the material pressure, $S_{\text{rp}, i}$ is the radiation-momentum source such that $- S_{\text{rp}, i}$ is a material-momentum source, $E = \tfrac{1}{2} \, \rho \, u^2 + \rho \, e$ and $E_{ij} \equiv \tfrac{1}{2} \, \rho \, u_i \, u_j + \rho \, e \, \delta_{ij}$ are defined for notational convenience, $S_{\text{re}}$ is the radiation-energy source such that $- S_{\text{re}}$ is a material-energy source, $c$ is the speed of light, and $\nu$ and $\nu_{\text{o}}$ are the LF and CMF frequencies, respectively.
We leave the Planck function in the CMF,
% \begin{linenomath}
\begin{gather}
  B_{\nu_{\text{o}}}(T) = \frac{2 \, h \, \nu_{\text{o}}^3}{c^2} \left[ e^{h \, \nu_{\text{o}} / k_{\text{B}} \, T} - 1 \right]^{-1} \, ,
\end{gather}
% \end{linenomath}
as a function of the CMF frequency, $\nu_{\text{o}}$, and the local material temperature, $T$, where $h$ is Planck's constant and $k_{\text{B}}$ is Boltzmann's constant.
Finally, $\sigma_{\text{s}}$, $\sigma_{\text{a}, \nu_{\text{o}}}$, and $\sigma_{\text{t}, \nu_{\text{o}}} = \sigma_{\text{a}, \nu_{\text{o}}} + \sigma_{\text{s}}$, are the scattering, absorption and total cross-sections in the CMF; $\sigma_{\text{t}, \nu_{\text{o}}}$ and $\sigma_{\text{a}, \nu_{\text{o}}}$ are functions of $\nu_{\text{o}}$ while $\sigma_{\text{s}}$ is independent of frequency.
We omit the subscript-o from the Planck function and the cross sections, which otherwise denotes a CMF variable.

The LF radiation-energy and radiation-momentum sources, $S_{\text{re}}$ and $S_{\text{rp}, i}$, are the first two frequency-integrated angular moments of $Q_{\nu}$, respectively:
\begin{subequations}
\label{eqs:radiation_sources}
% \begin{linenomath}
  \begin{gather}
  \label{eq:LF_Sre}
    S_{\text{re}} \equiv \int_{4 \pi} \int_0^{\infty} Q_{\nu} \, d\Omega \, d\nu = \partial_t {\cal E} + \partial_i {\cal F}_i \, ,
  \end{gather}
% \end{linenomath}
  \vspace{-20pt}
% \begin{linenomath}
  \begin{multline}
  \label{eq:LF_Srp}
    S_{\text{rp}, i} \equiv \frac{1}{c} \int_{4 \pi} \int_0^{\infty} \Omega_i \, Q_{\nu} \, d\Omega \, d\nu \\
    = \frac{1}{c^2} \partial_t {\cal F}_i + \partial_j {\cal P}_{ij} \, ,
  \end{multline}
% \end{linenomath}
\end{subequations}
and
\begin{subequations}
\label{eqs:LF_radiation_variables}
% \begin{linenomath}
  \begin{gather}
    {\cal E} \equiv \frac{1}{c} \int_{4 \pi} \int_0^{\infty} I_{\nu} \, d\Omega \, d\nu \, , \\
    {\cal F}_i \equiv \int_{4 \pi} \int_0^{\infty} \Omega_i \, I_{\nu} \, d\Omega \, d\nu \, , \\
    {\cal P}_{ij} \equiv \frac{1}{c} \int_{4 \pi} \int_0^{\infty} \Omega_i \, \Omega_j \, I_{\nu} \, d\Omega \, d\nu \, ,
  \end{gather}
% \end{linenomath}
\end{subequations}
are the LF radiation energy density, radiation flux, and radiation-pressure, which are the first three frequency-integrated angular moments of $I_{\nu}$, respectively.
The angle- and frequency-integrals of the radiation sources (\ref{eqs:radiation_sources}) cannot be analytically performed
% without introducing simplifying approximations
since $Q_{\nu}$ (\ref{eq:LF_RT_source}) contains angle- and frequency-dependent variables whose functional form is not generally known.
\subsection{The Equilibrium-diffusion approximation}
\label{subsec:CMF_EDA}
The EDA imposes four basic assumptions on a physical system \cite{MM1999}:
1) the photon mean-free-path is small compared to the size of the absorption-dominated system,
2) the matter-radiation system is in thermal equilibrium,
\begin{comment}
  JEM: See previous comment. (about diffusion in the introduction)
\end{comment}
% 3) the radiation flux is diffusive and time-independent, and
3) the radiation flux is diffusive, and
4) the radiation pressure is isotropic.
These simplifications allow for a basic understanding of the underlying phenomena, the equations describing them, and the form their solutions might take.
In the LF the EDA and its simplified set of RH equations are:
\begin{subequations}
\label{eq:LF_EDA_radiation_variables}
% \begin{linenomath}
  \begin{gather}
    {\cal E} = a_{\text{\tiny R}} \, T^4 \, , \label{eq:LF_RED} \\
    {\cal F}_i = - \frac{a_{\text{\tiny R}} \, c}{3 \, \sigma_{\text{t}, \text{\tiny R}}} \, \partial_i T^4 + \frac{4}{3} \, u_i \, a_{\text{\tiny R}} \, T^4 \, , \label{eq:LF_Fr} \\
    {\cal P}_{ij} = \frac{1}{3} \, a_{\text{\tiny R}} \, T^4 \, \delta_{ij} \, , \label{eq:LF_Pr}
  \end{gather}
% \end{linenomath}
\end{subequations}
\vspace{-25pt}
\begin{subequations}
\label{eqs:EDA_RH_equations}
% \begin{linenomath}
  \begin{gather}
    \partial_t \rho + \partial_i \left( \rho \, u_i \right) = 0 \, , \label{eq:EDA_mass_conservation}
  \end{gather}
  \vspace{-25pt}
  \begin{multline}
    \partial_t \left( \rho \, u_i \right) \\
    + \partial_j \left( \rho \, u_i \, u_j + p_{ij} + \frac{1}{3} \, a_{\text{\tiny R}} \, T^4 \, \delta_{ij} \right) = 0 \, , \label{eq:EDA_momentum_conservation}
  \end{multline}
  \vspace{-20pt}
  \begin{multline}
    \partial_t \left( E + {\cal E} \right) \\
    + \partial_i \left[ u_j \left( E_{ij} + p_{ij} + \frac{4}{3} \, a_{\text{\tiny R}} \, T^4 \, \delta_{ij} \right) \right] = \\
    \partial_i \left( \frac{a_{\text{\tiny R}} \, c}{3 \, \sigma_{\text{t}, \text{\tiny R}}} \, \partial_i T^4 \right) \, , \label{eq:EDA_energy_conservation}
  \end{multline}
% \end{linenomath}
\end{subequations}
where $a_{\text{\tiny R}}$ is the radiation constant and $\sigma_{\text{t}, \text{\tiny R}}$ is the CMF Rosseland-averaged cross section \cite{MM1999}.
We point out that the time-derivative of the radiation flux in the total-momentum equation (\ref{eq:EDA_momentum_conservation}) has been dropped in accordance with the diffusion approximation.

The strength of the EDA in solving RH problems is that the radiation variables (\ref{eq:LF_EDA_radiation_variables}) in this simplified set of equations (\ref{eqs:EDA_RH_equations}) are explicit functions of the hydrodynamic variables.
Thus, once an equation-of-state for the material is specified, these simplified RH equations represent a solvable system that are easier to solve than the full set of RH equations (\ref{eqs:RH_equations}) \cite{LM2001}.
To be clear, the EDA neglects most of the radiation information that is contained in the RT equation (\ref{eq:LF_RT}), and this is why we refer to any terms beyond the EDA as transport corrections.
\section{Asymptotic analysis of the RH equations}
\label{sec:asymptotic_analysis}
In this section, we present our main result which is that the EDA (\ref{eq:LF_EDA_radiation_variables}) and its simplified set of RH equations (\ref{eqs:EDA_RH_equations}) are first-order accurate in the EDL with transport corrections beginning at second order.
These results are derived in \ref{app:app3A}.
In Subsection \ref{subsec:nondimensionalization}, the RH variables and simplified RH equations are nondimensionalized and nondimensional parameters are defined.
The EDL scalings were originally presented by Lowrie, Morel and Hittinger \cite{LMH1999}, and we use these in Subsection \ref{subsec:asymptotic_EDL_scalings} to scale the nondimensional RH equations.
Finally, the scaled equations are asymptotically expanded and the main result of this paper is presented in Subsection \ref{subsec:asymptotic_EDL_results}.
This analysis and its results only apply to the interior portion of a RH system, and do not account for effects due to initial nor boundary conditions, nor boundary layers.
\subsection{Nondimensionalization of the RH equations}
\label{subsec:nondimensionalization}
\begin{comment}
  JEM: I would indicate that the dimensional variable is constant.
       It would work to insert "constant" between "a" and "dimensional".
  JMF: Fixed
\end{comment}
\begin{comment}
  JEM: dimensionality makes me think of the number of dimensions rather than the dimensions themselves.  Perhaps it would be clearer to replace "dimensionality" with "dimensions".
  JMF: Fixed.
\end{comment}
Each dimensional variable is decomposed into the product of a constant dimensional variable with subscript-$\infty$, indicating a reference state, which only retains the
%dimensionality
dimension of the original variable, and a non-constant nondimensional variable with a hat, which retains the value of the original variable:
% \begin{linenomath}
\begin{gather*}
  x = \hat{x} \, l_{\infty} \, , \quad t = \hat{t} \, \frac{l_{\infty}}{a_{\infty}} \, , \quad u = \hat{u} \, u_{\infty} \, , \\
  \rho = \hat{\rho} \, \rho_{\infty} \, , \quad p = \hat{p} \, \rho_{\infty} \, a_{\infty}^2 \, , \quad e = \hat{e} \, a_{\infty}^2 \, , \\
  T = \hat{T} \, T_{\infty} \, , \quad \sigma_{\text{t}} = \hat{\sigma}_{\text{t}} \, \sigma_{\text{t}, \infty} \, , \quad \sigma_{\text{s}} = \hat{\sigma}_{\text{s}} \, \sigma_{\text{s}, \infty} \, , \\
  I_{\nu} = \hat{I}_{\hat{\nu}} \, \frac{a_{\text{\tiny R}} \, c \, h \, T_{\infty}^3}{k_{\text{B}}} \, , \quad \nu = \hat{\nu} \, \frac{k_{\text{B}} \, T_{\infty}}{h} \, ,
\end{gather*}
% \end{linenomath}
where $l_{\infty}$ is the reference length of the system, $a_{\infty}$ is a reference sound speed for the fluid, $u_{\infty}$ is a reference fluid velocity, $\rho_{\infty}$ is a reference fluid mass density, and $T_{\infty}$ is a reference fluid temperature.
The total and scattering reference cross sections are $\sigma_{\text{t}, \infty}$ and $\sigma_{\text{s}, \infty}$, respectively.
The radiation intensity and frequency, $\hat{I}_{\hat{\nu}}$ and $\hat{\nu}$,
% respectively,
are nondimensionalized in such a way that the nondimensional radiation energy density, radiation flux, and radiation pressure are consistent with their definitions (\ref{eqs:LF_radiation_variables}) above:
\begin{subequations}
% \begin{linenomath}
  \begin{gather}
    {\cal E} = \hat{{\cal E}} \, a_{\text{\tiny R}} \, T_{\infty}^4 \, , \\
    {\cal F}_i = \hat{{\cal F}}_i \, a_{\text{\tiny R}} \, c \, \, T_{\infty}^4 \, , \\
    {\cal P}_{ij} = \hat{{\cal P}}_{ij} \, a_{\text{\tiny R}} \, T_{\infty}^4 \, .
  \end{gather}
% \end{linenomath}
\end{subequations}
The nondimensionalized form of the Euler equations coupled to the radiation energy and momentum sources
% , equations (\ref{eq:mass_conservation}) - (\ref{eq:energy_conservation}),
are:
\begin{subequations}
\label{eq:nondimensional_RH_conservation}
% \begin{linenomath}
  \begin{gather}
    \partial_t \rho + {\cal M} \, \partial_i \left( \rho \, u_i \right) = 0 \, ,
  \end{gather}
% \end{linenomath}
  \vspace{-30pt}
% \begin{linenomath}
  \begin{multline}
    {\cal M} \, \partial_t \left( \rho \, u_i \right) + \partial_j \left( {\cal M}^2 \, \rho \, u_i \, u_j + p_{ij} \right) = \\
    - P_0 \left( {\cal U} \, \partial_t {\cal F}_i + \partial_j {\cal P}_{ij} \right) \, ,
  \end{multline}
% \end{linenomath}
  \vspace{-25pt}
% \begin{linenomath}
  \begin{multline}
    \partial_t \left( \frac{1}{2} \, {\cal M}^2 \, \rho \, u^2 + \rho \, e \right) \\
    + \partial_i \left[ {\cal M} \, u_j \left( \frac{1}{2} \, {\cal M}^2 \, \rho \, u_i \, u_j + \rho \, e \, \delta_{ij} + p_{ij} \right) \right] \\
    = - P_0 \left( \partial_t {\cal E} + \, {\cal C} \, \partial_i {\cal F}_i \right) \, ,
  \end{multline}
% \end{linenomath}
\end{subequations}
where the hats have been dropped for notational convenience.
The nondimensional parameters are:
\begin{subequations}
% \begin{linenomath}
  \begin{gather*}
    {\cal M} \equiv \frac{u_{\infty}}{a_{\infty}} \, , \quad P_0 \equiv \frac{a_{\text{\tiny R}} \, T_{\infty}^4}{\rho_{\infty} \, a_{\infty}^2} \, , \\
    {\cal U} \equiv \frac{a_{\infty}}{c} \, , \quad {\cal C} \equiv \frac{c}{a_{\infty}} \, , \\
    {\cal L} \equiv \frac{l_{\infty}}{\lambda_{\text{t}, \infty}} = l_{\infty} \, \sigma_{\text{t}, \infty} \, , \quad {\cal L}_{\text{s}} \equiv \frac{\sigma_{\text{s}, \infty}}{\sigma_{\text{t}, \infty}} \, ,
  \end{gather*}
% \end{linenomath}
\end{subequations}
the first four of which are used in equations (\ref{eq:nondimensional_RH_conservation}) and the last two are from \ref{app:app3A}; ${\cal M}$ is related to the Mach number of the material flow, $P_0$ is a measure of the influence radiation has on the material, ${\cal U}$ is the ratio of the reference sound speed to the speed of light, ${\cal C}$ is the inverse of ${\cal U}$, ${\cal L}$ is a measure of the system's size compared to the radiation mean-free-path, and ${\cal L}_{\text{s}}$ is a measure of whether the system is absorption or scattering dominated.
\subsection{Asymptotic scalings}
\label{subsec:asymptotic_EDL_scalings}
The EDL scalings of the nondimensional parameters were originally introduced by Lowrie, Morel and Hittinger \cite{LMH1999}:
\begin{subequations}
% \begin{linenomath}
  \begin{gather*}
    {\cal M} = {\cal O}(1) \, , \quad P_0 = {\cal O}(1) \, , \\
    {\cal U} = {\cal O}(\epsilon) \, , \quad {\cal C} = {\cal O}(\epsilon^{-1}) \, , \\
    {\cal L} = {\cal O}(\epsilon^{-1}) \, , \quad {\cal L}_{\text{s}} = {\cal O}(\epsilon) \, .
  \end{gather*}
% \end{linenomath}
\end{subequations}
The first scaling implies that no assumption is made about whether the value of ${\cal M}$ should be small nor large.
As such, the EDL supports shock-wave solutions as shown by Lowrie and Rauenzahn \cite{LR2007}, which is a result we will use in Subsection \ref{subsec:LF_nEDA_scales_to_EDA}.
% The second scaling similarly implies that no assumption is made about whether the amount of radiation in the system is small nor large, so it may be negligible or it may dominate the system.
The second scaling similarly implies that no assumption is made about whether the system is radiation dominated or not.
The third and fourth scalings are consistent with nonrelativistic hydrodynamics.
The fifth scaling implies that the reference length of the system is much larger than the photon mean-free-path, 
% such that it is reasonable to assume that radiation diffuses through the system.
so that radiation may diffuse through the system.
The sixth scaling implies that absorption dominates scattering.
These last two scalings are used
% in the RT equation
in \ref{app:app3A}.
Reviewing the nondimensionalized equations (\ref{eq:nondimensional_RH_conservation})
% , with these scalings in mind,
we see that only the radiation flux is affected by the scalings.
% Using this in equations (\ref{eq:nondimensional_RH_conservation}) produces:
The redimensionalized and scaled version of equations (\ref{eq:nondimensional_RH_conservation}) is:
\begin{subequations}
\label{eq:RH_scaled}
% \begin{linenomath}
  \begin{gather}
    \partial_t \rho + \partial_i \left( \rho \, u_i \right) = 0 \, , \label{eq:mass_conservation_scaled}
  \end{gather}
% \end{linenomath}
  \vspace{-25pt}
% \begin{linenomath}
  \begin{multline}
    \partial_t \left( \rho \, u_i \right) + \partial_j \left( \rho \, u_i \, u_j + p_{ij} + {\cal P}_{ij} \right) = \\
    - \epsilon \, \frac{1}{c^2} \, \partial_t {\cal F}_i \, , \label{eq:momentum_conservation_scaled}
  \end{multline}
% \end{linenomath}
  \vspace{-15pt}
% \begin{linenomath}
  \begin{multline}
    \partial_t \left( E + {\cal E} \right) + \partial_i \left[ u_j \left( E_{ij} + p_{ij} \right) \right] = - \frac{1}{\epsilon} \, \partial_i {\cal F}_i \, , \label{eq:energy_conservation_scaled}
  \end{multline}
% \end{linenomath}
\end{subequations}
% which have been re-dimensionalized here for notational convenience, and
where we have moved all terms without an $\epsilon$ to the left-hand side.
No approximations have gone into these equations nor any of their variables yet.
\subsection{Asymptotic expansion and results}
\label{subsec:asymptotic_EDL_results}
The Euler equations are first-order accurate using scalings that agree with those above \cite{CC1958}.
The material and radiation variables in equations (\ref{eq:RH_scaled}) are now expanded in powers of $\epsilon$; as an example, the expansion for the material mass density is:
% \begin{linenomath}
\begin{gather}
  \rho = \sum_{n = 0}^{\infty} \rho^{(n)} \, \epsilon^n \, .
\end{gather}
% \end{linenomath}
Collecting equations at equal orders in $\epsilon$ forms an infinite hierarchical set.
The zeroth- and first-order contributions from equations (\ref{eq:RH_scaled}) are:%  \\
% $\underline{{\cal O}(\epsilon^0)}$
\begin{subequations}
\label{eq:scaled_conservation_equations}
% \begin{linenomath}
  \begin{gather}
    \left[ \partial_t \rho + \partial_i \left( \rho \, u_i \right) \right]^{(0)} = 0 \, , \\
    \left[ \partial_t \left( \rho \, u_i \right) + \partial_j \left( \rho \, u_i \, u_j + p_{ij} + {\cal P}_{ij} \right) \right]^{(0)} = 0 \, ,
  \end{gather}
% \end{linenomath}
  \vspace{-30pt}
% \begin{linenomath}
  \begin{multline}
    \left\{ \partial_t \left( E + {\cal E} \right) + \partial_i \left[ u_j \left( E_{ij} + p_{ij} \right) \right] \right\}^{(0)} \\
    = - \partial_i {\cal F}_i^{(1)} \, ,
  \end{multline}
% \end{linenomath}
% $\underline{{\cal O}(\epsilon^1)}$
% \begin{linenomath}
  \begin{gather}
    \left[ \partial_t \rho + \partial_i \left( \rho \, u_i \right) \right]^{(1)} = 0 \, ,
  \end{gather}
% \end{linenomath}
  \vspace{-30pt}
% \begin{linenomath}
  \begin{multline}
    \left[ \partial_t \left( \rho \, u_i \right) + \partial_j \left( \rho \, u_i \, u_j + p_{ij} + {\cal P}_{ij} \right) \right]^{(1)} \\
    = - \frac{1}{c^2} \, \partial_t {\cal F}_i^{(0)} \, ,
  \end{multline}
% \end{linenomath}
  \vspace{-30pt}
% \begin{linenomath}
  \begin{multline}
    \left\{ \partial_t \left( E + {\cal E} \right) + \partial_i \left[ u_j \left( E_{ij} + p_{ij} \right) \right] \right\}^{(1)} \\
    = - \partial_i {\cal F}_i^{(2)} \, .
  \end{multline}
% \end{linenomath}
\end{subequations}
The radiation variables ${\cal E}^{(0)}$, ${\cal F}_i^{(0)}$, ${\cal P}_{ij}^{(0)}$, ${\cal E}^{(1)}$, ${\cal F}_i^{(1)}$, ${\cal P}_{ij}^{(1)}$, and ${\cal F}_i^{(2)}$ are determined by analyzing the RT equation.
This is done in \ref{app:app3A} and we present the results here:%  \\
% \begin{linenomath}
\begin{subequations}
\label{eq:asymptotic_RH_results}
% $\underline{{\cal O}(\epsilon^0)}$
  \begin{gather}
    {\cal E}^{(0)} = \left[ a_{\text{\tiny R}} \, T^4 \right]^{(0)} \, , \\
    {\cal F}_i^{(0)} = 0 \, , \label{eq:F0} \\
    {\cal P}_{ij}^{(0)} = \left[ \frac{1}{3} \, a_{\text{\tiny R}} \, T^4 \, \delta_{ij} \right]^{(0)} \, ,
  \end{gather}
% $\underline{{\cal O}(\epsilon^1)}$
  \begin{gather}
    {\cal E}^{(1)} = \left[ a_{\text{\tiny R}} \, T^4 \right]^{(1)} \, , \\
    {\cal F}_i^{(1)} = \left[ - \frac{a_{\text{\tiny R}} \, c}{3 \, \sigma_{\text{t}, \text{\tiny R}}} \, \partial_i T^4 + \frac{4}{3} \, u_i \, a_{\text{\tiny R}} \, T^4 \right]^{(0)} \, , \\
    {\cal P}_{ij}^{(1)} = \left[ \frac{1}{3} \, a_{\text{\tiny R}} \, T^4 \, \delta_{ij} \right]^{(1)} \, ,
  \end{gather}
% $\underline{{\cal O}(\epsilon^2)}$
  \begin{gather}
    {\cal F}_i^{(2)} = \left[ - \frac{a_{\text{\tiny R}} \, c}{3 \, \sigma_{\text{t}, \text{\tiny R}}} \, \partial_i T^4 + \frac{4}{3} \, u_i \, a_{\text{\tiny R}} \, T^4 \right]^{(1)} \, .
  \end{gather}
\end{subequations}
% \end{linenomath}
Summing these variables through first order produces the EDA as given in equations (\ref{eq:LF_EDA_radiation_variables}):
\begin{subequations}
\label{eq:asymptotic_equilibrium_diffusion_solutions}
% \begin{linenomath}
  \begin{gather}
    {\cal E} = a_{\text{\tiny R}} \, T^4 \, , \label{eq:radiation_energy_density_order1} \\
    {\cal F}_i = - \frac{a_{\text{\tiny R}} \, c}{3 \, \sigma_{\text{t}, \text{\tiny R}}} \, \partial_i T^4 + \frac{4}{3} \, u_i \, a_{\text{\tiny R}} \, T^4 \, , \label{eq:radiation_flux_order2} \\
    {\cal P}_{ij} = \frac{1}{3} \, a_{\text{\tiny R}} \, T^4 \, \delta_{ij} \, , \label{eq:radiation_pressure_order1}
  \end{gather}
% \end{linenomath}
\end{subequations}
which are in agreement with the results presented in \cite{LMH1999}.
The derivation in \ref{app:app3A} shows that the radiation energy density and radiation pressure contain transport corrections to the EDA at second order, and so the EDA is at most first-order accurate in the EDL.
Summing the scaled equations (\ref{eq:scaled_conservation_equations}) through first order, and using the results just presented (\ref{eq:asymptotic_RH_results}), produces the EDA's simplified RH equations (\ref{eqs:EDA_RH_equations}):
\begin{subequations}
\label{eq:first_order_conservation_equations}
% \begin{linenomath}
  \begin{gather}
    \partial_t \rho + \partial_i \left( \rho \, u_i \right) = 0 \, ,
  \end{gather}
% \end{linenomath}
  \vspace{-25pt}
% \begin{linenomath}
  \begin{multline}
     \partial_t \left( \rho \, u_i \right) \\
    + \partial_j \left( \rho \, u_i \, u_j + p_{ij} + \frac{1}{3} \, a_{\text{\tiny R}} \, T^4 \, \delta_{ij} \right) = 0 \, ,
  \end{multline}
% \end{linenomath}
  \vspace{-20pt}
% \begin{linenomath}
  \begin{multline}
    \partial_t \left( E + a_{\text{\tiny R}} \, T^4 \right) \\
    + \partial_i \left[ u_j \left( E_{ij} + p_{ij} + \frac{4}{3} \, a_{\text{\tiny R}} \, T^4 \, \delta_{ij} \right) \right] \\
    = \partial_i \left( \frac{a_{\text{\tiny R}} \, c}{3 \, \sigma_{\text{t}, \text{\tiny R}}} \, \partial_i T^4 \right) \, . \label{eq:first_order_energy_conservation}
  \end{multline}
% \end{linenomath}
\end{subequations}
Therefore, the EDA and its simplified RH equations are first-order accurate with transport corrections occurring at second order.
This is the main result of this paper.
Both simplified models and numerical discretizations of the RH equations should preserve this asymptotic behavior.
Models that have different asymptotic behavior may not attain the correct solution in or near the EDL, while improper discretizations may require an unreasonable computational cost.
These results only apply to the interior solution of an RH problem far from any boundaries, and sufficiently late in time, so that the initial and boundary conditions may be neglected.
\section{Analysis of two grey RH approximations}
\label{sec:LF_CMF}
We have established the first-order accuracy of the EDA and its simplified RH equations.
As a reminder, the assumptions comprising the EDA are listed at the beginning of Subsection \ref{subsec:CMF_EDA}.
In this section, we analyze two RH approximations that are commonly used to solve a wide span of RH problems and we show that they both preserve the EDA's first-order accuracy.
The two approximations are the frequency-independent (``grey'') nonequilibrium-diffusion approximation and the grey Eddington approximation.
These approximations are applied to the radiation energy and momentum sources which are presented in Subsection \ref{subsec:grey_radiation_sources}.
The grey nonequilibrium-diffusion approximation is analyzed in Subsection \ref{subsec:CMF_grey-diffusion}.
Compared to the EDA it relaxes the requirement that the radiation be in thermal equilibrium with the material, hence the moniker ``nonequilibrium''.
As a practical example, in Subsection \ref{subsec:LF_nEDA_scales_to_EDA}, we present a radiative-shock solution from this approximation and show that the EDA solution is obtained when $\epsilon$ is small.
In contrast to the nonequilibrium-diffusion approximation, the grey Eddington approximation only retains the EDA assumption that the radiation pressure is isotropic.
As such, it can be used to describe environments where diffusion does not apply, or that may be out of equilibrium, or which may be highly rarefied.
This approximation is analyzed in Subsection \ref{subsec:CMF_grey-Eddington}.
\subsection{The grey radiation sources}
\label{subsec:grey_radiation_sources}
The two approximations analyzed in this section are typically made in the CMF by modifying the radiation energy and momentum sources.
These sources are equations (6.47) and (6.48) in \cite{Castor2007}:
\begin{subequations}
\label{eq:comoving_frame_source_rate_equations}
% \begin{linenomath}
  \begin{multline}
    \partial_t {\cal E}_{\text{o}} + \frac{1}{c} \, \partial_t \left( \beta_i \, {\cal F}_{\text{o}, i} \right) + \frac{a_i}{c^2} \, {\cal F}_{\text{o}, i} \\
    + \partial_i {\cal F}_{\text{o}, i} + \partial_i \left( u_i \, {\cal E}_{\text{o}} \right) + {\cal P}_{\text{o}, ij} \, \partial_j u_i \\
    = \sigma_{\text{a}} \, c \left( a_{\text{\tiny R}} \, T^4 - {\cal E}_{\text{o}} \right) \, , \label{subeq:CMF_Sre}
  \end{multline}
  \vspace{-25pt}
  \begin{multline}
    \frac{1}{c^2} \, \partial_t {\cal F}_{\text{o}, i} + \frac{a_i}{c^2} \, {\cal E}_{\text{o}} + \frac{1}{c} \, \partial_t \left( \beta_j \, {\cal P}_{\text{o}, ij} \right) \\
    + \partial_j {\cal P}_{\text{o}, ij} + \frac{1}{c} \, {\cal F}_{\text{o}, j} \, \partial_j \beta_i + \frac{1}{c} \, \partial_j \left( \beta_j \, {\cal F}_{\text{o}, i} \right) \\
    = - \frac{\sigma_{\text{t}}}{c} \, {\cal F}_{\text{o}, i} \, , \label{subeq:CMF_Srp}
  \end{multline}
% \end{linenomath}
\end{subequations}
where we have dropped the ${\cal O}(\beta^2)$ terms coming from the Lagrangian derivatives.
In order to compare the CMF approximation to our results we Lorentz transform the approximation to the LF and analyze it there.
A discussion of how to perform Lorentz transformations is outside the scope of this paper and can be found in \cite{MM1999}.
Although the Lorentz transformation changes the coordinate frame being considered it does not alter the physical content of the approximation.
It should also be clear that it does not matter if the CMF radiation sources are scaled first or transformed first, so long as the scalings are also applied to the transformation.
Finally, in order to facilitate comparison of the radiation sources from the transformed approximation with the LF sources we record here the LF radiation energy and momentum sources:
\begin{subequations}
\label{eq:lab_frame_source_rate_equations}
% \begin{linenomath}
  \begin{multline}
    \partial_t {\cal E} + \partial_i {\cal F}_i = \sigma_{\text{a}} \, c \left( a_{\text{\tiny R}} \, T^4 - {\cal E} \right) \\
    + \beta_i \left( \sigma_{\text{a}} - \sigma_{\text{s}} \right) {\cal F}_i \, ,
  \end{multline}
  \vspace{-25pt}
  \begin{multline}
    \frac{1}{c^2} \, \partial_t {\cal F}_i + \partial_j {\cal P}_{ij} = - \frac{\sigma_{\text{t}}}{c} \, {\cal F}_i \\
    + \beta_j \left( \sigma_{\text{t}} \, {\cal P}_{ij} + \sigma_{\text{s}} \, {\cal E} \, \delta_{ij} + \sigma_{\text{a}} \, a_{\text{\tiny R}} \, T^4 \, \delta_{ij} \right) \, ,
  \end{multline}
% \end{linenomath}
\end{subequations}
which are equations (2.29) and (2.30) in \cite{MK1982}.
The simplest way to produce the scaled radiation-source equations is to make the following replacements,
% \begin{linenomath}
\begin{gather*}
  \partial_t \rightarrow \epsilon^2 \, \partial_t \, , \quad \partial_i \rightarrow \epsilon \, \partial_i \, , \quad \beta_i \rightarrow \epsilon \, \beta_i \, , \quad \sigma_{\text{s}} \rightarrow \epsilon \, \sigma_{\text{s}} \, ,
\end{gather*}
% \end{linenomath}
and then to divide the radiation-energy source by $\epsilon^2$ and the radiation-momentum source by $\epsilon$.
The reason for dividing the radiation sources by $\epsilon$ and $\epsilon^2$ is to keep these sources at the corresponding order of their associated hydrodynamic sources.
\subsection{The nonequilibrium-diffusion approximation}
\label{subsec:CMF_grey-diffusion}
The nonequilibrium-diffusion approximation modifies the radiation-source equations by imposing the Eddington approximation, ${\cal P}_{\text{o}, ij} = \tfrac{1}{3} \, {\cal E}_{\text{o}} \, \delta_{ij}$ or ${\cal P}_{\text{o}, ii} = {\cal E}$, dropping the time-derivative of the radiation flux and all acceleration terms, and dropping all other terms in the radiation-momentum source until Fick's First Law of Diffusion is obtained.
Applied to the CMF radiation sources (\ref{eq:comoving_frame_source_rate_equations}) it produces the following simplified source equations:
\begin{subequations}
\label{eqs:grey_CMF_nED_radiation_sources}
% \begin{linenomath}
  \begin{multline}
    \partial_t {\cal E}_{\text{o}} + \partial_i {\cal F}_{\text{o}, i} + \partial_i \left( u_i \, {\cal E}_{\text{o}} \right) + \frac{1}{3} \, {\cal E}_{\text{o}} \, \partial_i u_i \\
  = \sigma_{\text{a}} \, c \left( a_{\text{\tiny R}} \, T^4 - {\cal E}_{\text{o}} \right) \, ,
  \end{multline}
  \vspace{-20pt}
  \begin{gather}
    \frac{1}{3} \, \partial_i {\cal E}_{\text{o}} = - \frac{\sigma_{\text{t}}}{c} \, {\cal F}_{\text{o}, i} \, .
  \end{gather}
% \end{linenomath}
\end{subequations}
Solving the radiation-momentum source for the radiation flux and plugging this into the radiation-energy source produces:
\begin{subequations}
% \begin{linenomath}
  \begin{gather}
    {\cal F}_{o, i} = - \frac{c}{3 \, \sigma_{\text{t}}} \, \partial_i {\cal E}_{\text{o}} \, ,
  \end{gather}
  \vspace{-25pt}
  \begin{multline}
    \partial_t {\cal E}_{\text{o}} - \partial_i \left( \frac{c}{3 \, \sigma_{\text{t}}} \, \partial_i {\cal E}_{\text{o}} \right) + \partial_i \left( u_i \, {\cal E}_{\text{o}} \right) + \frac{1}{3} \, {\cal E}_{\text{o}} \, \partial_i u_i \\
  = \sigma_{\text{a}} \, c \left( a_{\text{\tiny R}} \, T^4 - {\cal E}_{\text{o}} \right) \, ,
  \end{multline}
% \end{linenomath}
\end{subequations}
which are equations (97.69) and (97.70) in \cite{MM1999}.
Lorentz transformation of the sources (\ref{eqs:grey_CMF_nED_radiation_sources}) along with the CMF Eddington approximation produces:
\begin{subequations}
% \begin{linenomath}
  \begin{multline}
    \partial_t {\cal E} + \partial_i {\cal F}_i - \frac{2}{c} \, \partial_t \left( \beta_i \, {\cal F}_i \right) \\
    = \sigma_{\text{a}} \, c \left( a_{\text{\tiny R}} \, T^4 - {\cal E} \right) + \beta_i \left( \sigma_{\text{a}} - \sigma_{\text{s}} \right) {\cal F}_i \, , \label{eq:CMF_nEDA_LF_energy_source}
  \end{multline}
% \end{linenomath}
  \vspace{-25pt}
% \begin{linenomath}
  \begin{multline}
    \frac{1}{3} \, \partial_i {\cal E} + \frac{\beta_i}{c} \, \partial_t {\cal E} + \frac{\beta_i}{c} \, \partial_j {\cal F}_j = \\
    - \frac{\sigma_{\text{t}}}{c} \, {\cal F}_i + \beta_i \left( \frac{1}{3} \, \sigma_{\text{t}} \, {\cal E} + \sigma_{\text{s}} \, {\cal E} + \sigma_{\text{a}} \, a_{\text{\tiny R}} \, T^4 \right) \, , \label{eq:CMF_nEDA_LF_momentum_source}
  \end{multline}
% \end{linenomath}
  \vspace{-25pt}
% \begin{linenomath}
  \begin{multline}
    {\cal P}_{ij} = \frac{1}{3} \, {\cal E} \, \delta_{ij} \\
    + \frac{1}{c} \left( \beta_j \, {\cal F}_i + \beta_i \, {\cal F}_j - \frac{2}{3} \, \beta_k \, {\cal F}_k \, \delta_{ij} \right) \, . \label{eq:CMF_nEDA_LF_radiation_pressure}
  \end{multline}
% \end{linenomath}
\end{subequations}
\begin{comment}
  JEM: I think we [should] explicitly mention that the CMF pressure transformed to the LF is not equal to the [LF] pressure, which is just equal to the first term in Eq. 21c.
  JMF: Added a sentence reiterating that the pressures are not equal.
\end{comment}
The ${\cal O}(\beta)$ term on the radiation pressure (\ref{eq:CMF_nEDA_LF_radiation_pressure}) is symmetric and traceless, so the Eddington approximation is retained after the Lorentz transformation, ${\cal P}_{ii} = {\cal E}$, since the trace of a traceless object is zero.
We quickly compare the transformed sources, (\ref{eq:CMF_nEDA_LF_energy_source}) and (\ref{eq:CMF_nEDA_LF_momentum_source}), with the LF sources (\ref{eq:lab_frame_source_rate_equations}).
If the nonequilibrium-diffusion approximation  were applied to the LF radiation sources (\ref{eq:lab_frame_source_rate_equations}) then the result would differ from equations (\ref{eq:CMF_nEDA_LF_energy_source}) and (\ref{eq:CMF_nEDA_LF_momentum_source}) by terms of ${\cal O}(\beta^2)$ and $\beta_i \, {\cal F}_i$, and the radiation pressure would neglect the symmetric traceless portion contained in equation (\ref{eq:CMF_nEDA_LF_radiation_pressure}).
% In the LF the Eddington approximation is ${\cal P}_{ii} = {\cal E}$, and so the radiation pressures compared in the LF are not equivalent.
% We will find that both of these terms are negligible in the EDL.
However, we will find that these differences between the CMF and LF sources, as well as the radiation pressures, are negligible in the EDL.
Scaling the radiation sources, (\ref{eq:CMF_nEDA_LF_energy_source}) and (\ref{eq:CMF_nEDA_LF_momentum_source}), gives:
\begin{subequations}
\label{eqs:scaled_CMF_nEDA_LF_radiation_sources}
% \begin{linenomath}
  \begin{multline}
    \partial_t {\cal E} + \frac{1}{\epsilon} \, \partial_i {\cal F}_i - \epsilon \, \frac{2}{c} \, \partial_t \left( \beta_i \, {\cal F}_i \right) \\
    = \frac{1}{\epsilon^2} \, \left( \sigma_{\text{t}} - \epsilon \, \sigma_{\text{s}} \right) c \left( a_{\text{\tiny R}} \, T^4 - {\cal E} \right) \\
    + \frac{1}{\epsilon} \, \beta_i \left( \sigma_{\text{t}} - 2 \, \epsilon \, \sigma_{\text{s}} \right) {\cal F}_i \, , \label{eq:scaled_CMF_nEDA_LF_energy_source}
  \end{multline}
% \end{linenomath}
  \vspace{-20pt}
% \begin{linenomath}
  \begin{multline}
    \frac{1}{3} \, \partial_i {\cal E} + \epsilon^2 \, \frac{\beta_i}{c} \, \partial_t {\cal E} + \epsilon \, \frac{\beta_i}{c} \, \partial_j {\cal F}_j \\
    = - \frac{1}{\epsilon} \, \frac{\sigma_{\text{t}}}{c} \, {\cal F}_i + \beta_i \Bigl( \frac{1}{3} \, \sigma_{\text{t}} \, {\cal E} + \epsilon \, \sigma_{\text{s}} \, {\cal E} \Bigr. \\
    \Bigl. + \left( \sigma_{\text{t}} - \epsilon \, \sigma_{\text{s}} \right) a_{\text{\tiny R}} \, T^4 \Bigr) \, . \label{eq:scaled_CMF_nEDA_LF_momentum_source}
  \end{multline}
% \end{linenomath}
\end{subequations}
The radiation-energy source (\ref{eq:scaled_CMF_nEDA_LF_energy_source}) provides the solution for the scaled radiation energy density,
\begin{subequations}
\label{eqs:CMF_nEDA_scaled_radiation_variables}
% \begin{linenomath}
  \begin{multline}
    {\cal E} = a_{\text{\tiny R}} \, T^4 \\
    - \epsilon \left[ \frac{\sigma_{\text{s}}}{\sigma_{\text{t}}} \left( a_{\text{\tiny R}} \, T^4 - {\cal E} \right) - \frac{\beta_i}{c} \, {\cal F}_i + \frac{1}{\sigma_{\text{t}} \, c} \, \partial_i {\cal F}_i \right] \, ,
  \end{multline}
% \end{linenomath}
where terms of ${\cal O}(\epsilon^2)$ have been discarded.
The radiation-momentum source (\ref{eq:scaled_CMF_nEDA_LF_momentum_source}) provides the solution for the scaled radiation flux,
% \begin{linenomath}
  \begin{multline}
    {\cal F}_i = \epsilon \left[ - \frac{c}{3 \, \sigma_{\text{t}}} \, \partial_i {\cal E} + u_i \left( \frac{1}{3} \, {\cal E} + a_{\text{\tiny R}} \, T^4 \right) \right] \\
    + \epsilon^2 \left[ - \frac{1}{\sigma_{\text{t}}} \, \beta_i \, \partial_j {\cal F}_j + u_i \, \frac{\sigma_{\text{s}}}{\sigma_{\text{t}}} \left( {\cal E} - a_{\text{\tiny R}} \, T^4 \right) \right] \, ,
  \end{multline}
% \end{linenomath}
where terms of ${\cal O}(\epsilon^3)$ have been discarded.
Scaling the radiation pressure (\ref{eq:CMF_nEDA_LF_radiation_pressure}) gives:
% \begin{linenomath}
  \begin{multline}
    {\cal P}_{ij} = \frac{1}{3} \, {\cal E} \, \delta_{ij} \\
    + \frac{1}{c} \, \epsilon \, \left( \beta_j \, {\cal F}_i + \beta_i \, {\cal F}_j - \frac{2}{3} \, \beta_k \, {\cal F}_k \, \delta_{ij} \right) \, .
  \end{multline}
% \end{linenomath}
\end{subequations}
From these expressions (\ref{eqs:CMF_nEDA_scaled_radiation_variables}) the zeroth- and first-order solutions are obtained along with the second-order radiation flux:
\begin{subequations}
\label{eqs:CMF_nEDA_LF_zeroth_first_radiation_variables}
% \begin{linenomath}
  \begin{gather}
    {\cal E}^{(0)} = \left[ a_{\text{\tiny R}} \, T^4 \right]^{(0)} \, , \\
    {\cal F}_i^{(0)} = 0 \, , \\
    {\cal P}_{ij}^{(0)} = \left[ \frac{1}{3} \, a_{\text{\tiny R}} \, T^4 \, \delta_{ij} \right]^{(0)} \, , \\
    {\cal E}^{(1)} = \left[ a_{\text{\tiny R}} \, T^4 \right]^{(1)} \, , \\
    {\cal F}_i^{(1)} = \epsilon \left[ - \frac{a_{\text{\tiny R}} \, c}{3 \, \sigma_{\text{t}}} \, \partial_i T^4 + \frac{4}{3} \, u_i \, a_{\text{\tiny R}} \, T^4 \right]^{(0)} \, , \\
    {\cal P}_{ij}^{(1)} = \left[ \frac{1}{3} \, a_{\text{\tiny R}} \, T^4 \, \delta_{ij} \right]^{(1)} \, , \\
    {\cal F}_i^{(2)} = \epsilon \left[ - \frac{a_{\text{\tiny R}} \, c}{3 \, \sigma_{\text{t}}} \, \partial_i T^4 + \frac{4}{3} \, u_i \, a_{\text{\tiny R}} \, T^4 \right]^{(1)} \, .
  \end{gather}
% \end{linenomath}
\end{subequations}
The zeroth- and first-order contributions from the scaled radiation sources (\ref{eqs:scaled_CMF_nEDA_LF_radiation_sources}) are:
\begin{subequations}
\label{eqs:CMF_nEDA_LF_zeroth_first_radiation_sources}
% \begin{linenomath}
  \begin{align}
    S_{\text{re}}^{(0)} & = \partial_t {\cal E}^{(0)} + \partial_i {\cal F}_i^{(1)} \, , \\
    S_{\text{rp}}^{(0)} & = \frac{1}{3} \, \partial_i {\cal E}^{(0)} \, , \\
    S_{\text{re}}^{(1)} & = \partial_t {\cal E}^{(1)} + \partial_i {\cal F}_i^{(2)} - \partial_t \left[ \frac{2}{c} \, \beta_i {\cal F}_i \right]^{(0)} \, , \\
    S_{\text{rp}}^{(1)} & = \frac{1}{3} \, \partial_i {\cal E}^{(1)} + \left[ \frac{\beta_i}{c} \, \partial_j {\cal F}_j \right]^{(0)} \, .
  \end{align}
% \end{linenomath}
\end{subequations}
Summing the zeroth- and first-order results for the radiation variables (\ref{eqs:CMF_nEDA_LF_zeroth_first_radiation_variables}) reproduces the EDA (\ref{eq:LF_EDA_radiation_variables}):
\begin{subequations}
% \begin{linenomath}
  \begin{gather}
    {\cal E} = a_{\text{\tiny R}} \, T^4 \, , \\
    {\cal F}_i = - \frac{a_{\text{\tiny R}} \, c}{3 \, \sigma_{\text{t}}} \, \partial_i T^4 + \frac{4}{3} \, u_i \, a_{\text{\tiny R}} \, T^4  \, , \\
    {\cal P}_{ij} = \frac{1}{3} \, a_{\text{\tiny R}} \, T^4 \, \delta_{ij} \, .
  \end{gather}
% \end{linenomath}
\end{subequations}
As a reminder, the radiation sources are coupled to the Euler equations; see equations (\ref{eqs:RH_equations}).
Summing the zeroth- and first-order contributions from the radiation sources (\ref{eqs:CMF_nEDA_LF_zeroth_first_radiation_sources}), coupled to the Euler equations (\ref{eqs:RH_equations}), and using the results in equations (\ref{eqs:CMF_nEDA_LF_zeroth_first_radiation_variables}), reproduces the EDA's simplified RH equations (\ref{eqs:EDA_RH_equations}):
\begin{subequations}
% \begin{linenomath}
  \begin{gather}
    \partial_t \rho + \partial_i \left( \rho u_i \right) = 0 \, ,
  \end{gather}
  \vspace{-25pt}
  \begin{multline}
    \partial_t \left( \rho \, u_i \right) \\
    + \partial_j \left( \rho \, u_i \, u_j + p_{ij} + \frac{1}{3} \, a_{\text{\tiny R}} \, T^4 \, \delta_{ij} \right) = 0 \, ,
  \end{multline}
  \vspace{-25pt}
  \begin{multline}
    \partial_t \left( E + a_{\text{\tiny R}} \, T^4 \right) \\
    + \partial_i \left[ u_j \left( E_{ij} + p_{ij} + \frac{4}{3} \, a_{\text{\tiny R}} \, T^4 \, \delta_{ij} \right) \right] \\
    = \partial_i \left( \frac{a_{\text{\tiny R}} \, c}{3 \, \sigma_{\text{t}}} \, \partial_i T^4 \right) \, .
  \end{multline}
% \end{linenomath}
\end{subequations}
Thus, the nonequilibrium-diffusion approximation preserves the EDA's first-order accuracy, and this result holds if the approximation is made in the CMF or the LF.
\begin{figure}[t!]
  \hspace{-15pt}
  \includegraphics{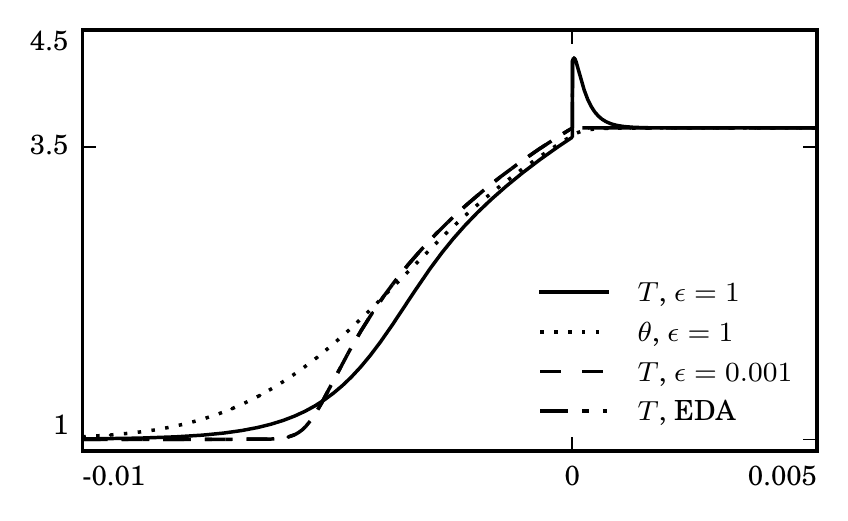}
\caption{
Comparison of the ${\cal M}_0 = 3$ nonequilibrium-diffusion radiative-shock solutions for $\epsilon = 1$ and $\epsilon = 0.001$.
These solutions were both produced using the solution method described by Lowrie and Edwards \cite{LE2008}, but with the inclusion of a factor of $\epsilon$, as described in Subsection \ref{subsec:LF_nEDA_scales_to_EDA}.
The solutions associated with $\epsilon = 1$ are the nonequilibrium-diffusion radiative-shock solutions, while the solution associated with $\epsilon = 0.001$ is the equilibrium-diffusion solution.
%; compare Figure 9 in \cite{LE2008}.
The equilibrium-diffusion solution produced by the method described by Lowrie and Rauenzahn is the dash-dotted line labeled as the EDA solution.
The EDA solution and the solution using $\epsilon = 0.001$ cannot be distinguished.
\label{fig:M3}}
\end{figure}
\subsection{Confirmation of the asymptotic analysis by radiative-shock solutions}
\label{subsec:LF_nEDA_scales_to_EDA}
In this subsection we show that a particular solution of the nonequilibrium-diffusion approximation transitions to the equilibrium-diffusion solution when $\epsilon$ is small.
Specifically, we briefly analyze the nonequilibrium-diffusion radiative-shock solution method developed by Lowrie and Edwards \cite{LE2008} and we show that when $\epsilon = 0.001$ that method produces the equilibrium-diffusion solution developed by Lowrie and Rauenzahn \cite{LR2007}.

The 1D nondimensional equations solved by Lowrie and Edwards, scaled in the EDL, are:
\begin{subequations}
\label{eqs:Lowrie_Edwards_scaled}
% \begin{linenomath}
  \begin{gather}
    \partial_x \left( \rho \, u \right) = 0 \, , \\
    \partial_x \left( \rho \, u^2 + p + \frac{1}{3} \, P_0 \, {\cal E} \right) = 0 \, , \\
    \partial_x \left[ u \left( \frac{1}{2} \, \rho \, u^2 + \rho \, e + p \right) \right] = - \frac{P_0}{\epsilon} \, \partial_x {\cal F} \, ,
  \end{gather}
% \end{linenomath}
  \vspace{-20pt}
% \begin{linenomath}
  \begin{multline}
    \frac{{\cal M}_0}{\gamma \left( \gamma - 1 \right)} \, \partial_x T + p \, \partial_x u = \\
    P_0 \, \sigma_{\text{t}} \left[ \frac{1}{\epsilon} \left( T^4 - {\cal E} \right) + 2 \, \beta \, {\cal F} \right] \, . \label{eq:LE_Srie}
  \end{multline}
% \end{linenomath}
\end{subequations}
The fourth equation (\ref{eq:LE_Srie}) is the radiation internal energy source, $S_{\text{rie}} = S_{\text{re}} - \beta \, S_{\text{rp}}$.
For a monatomic ideal-gas the adiabatic index is $\gamma = 5/3$, and ${\cal M}_0$ represents the initial Mach number of the unshocked gas, which we set to be ${\cal M}_0 = 3$ in this example.
The nondimensional radiation flux for a purely absorbing system is
% \begin{linenomath}
  \begin{gather}
    {\cal F} = - \frac{1}{3 \, \sigma_{\text{t}}} \, \partial_x {\cal E} + \frac{1}{3} \, \beta \, \sigma_{\text{t}} \left( {\cal E} + T^4 \right) \, .
  \end{gather}
% \end{linenomath}
When $\epsilon = 1$ we obtain the nonequilibrium-diffusion solution and when $\epsilon = 0.001$ we obtain the equilibrium-diffusion solution from the same
% set of equations and
solution method; see Figure \ref{fig:M3}.
% which can be compared with Figure 9 in \cite{LE2008}.
For comparison within the figure, we also include the equilibrium-diffusion solution produced by the method described by Lowrie and Rauenzahn, which is the dash-dotted line labeled as the EDA solution.
This EDA solution and the nonequilibrium-diffusion solution using $\epsilon = 0.001$ cannot be distinguished.
Thus, these shock problems can serve as a test problem to ensure that codes are asymptotic preserving.
However, other test problems are also needed that are relevant to different portions of the RH problem space.
\begin{comment}
  JEM: Do we gain anything by showing theta only for epsilon=1?
  JMF: We do not gain anything by adding another curve that is indistinguishable from the $T, \epsilon = 0.001$ curve and the $T, \text{EDA}$ curves.
\end{comment}
\subsection{The Eddington approximation}
\label{subsec:CMF_grey-Eddington}
In this subsection we analyze the grey Eddington approximation applied to the CMF radiation sources (\ref{eq:comoving_frame_source_rate_equations}).
% This approximation modifies the radiation sources by defining the radiation pressure as ${\cal P}_{\text{o}, ij} = \tfrac{1}{3} \, {\cal E}_{\text{o}} \, \delta_{ij}$.
The application of this approximation produces the following CMF radiation sources:
\begin{subequations}
% \begin{linenomath}
  \begin{multline}
    \partial_t {\cal E}_{\text{o}} + \frac{1}{c} \partial_t \left( \beta_i \, {\cal F}_{\text{o}, i} \right) + \frac{a_i}{c^2} {\cal F}_{\text{o}, i} \\
    + \partial_i {\cal F}_{\text{o}, i} + \partial_i \left( u_i \, {\cal E}_{\text{o}} \right) + \frac{1}{3} \, {\cal E}_{\text{o}} \, \partial_i u_i \\
    = \sigma_{\text{a}} \, c \left( a_{\text{\tiny R}} \, T^4 - {\cal E}_{\text{o}} \right) \, ,
  \end{multline}
  \vspace{-25pt}
  \begin{multline}
    \frac{1}{c^2} \partial_t {\cal F}_{\text{o}, i} + \frac{a_i}{c^2} {\cal E}_{\text{o}} + \frac{1}{3 \, c} \partial_t \left( \beta_i \, {\cal E}_{\text{o}} \right) \\
    + \frac{1}{3} \, \partial_i {\cal E}_{\text{o}} + \frac{1}{c} {\cal F}_{\text{o}, j} \, \partial_j \beta_i + \frac{1}{c} \partial_j \left( \beta_j \, {\cal F}_{\text{o}, i} \right) \\
    = - \frac{\sigma_{\text{t}}}{c} {\cal F}_{\text{o}, i} \, .
  \end{multline}
% \end{linenomath}
\end{subequations}
Lorentz transformation of these sources along with the Eddington approximation to the LF produces:
\begin{subequations}
\label{eqs:LF_grey-Eddington_sources}
% \begin{linenomath}
  \begin{multline}
    \partial_t {\cal E} + \partial_i {\cal F}_i = \sigma_{\text{a}} \, c \left( a_{\text{\tiny R}} \, T^4 - {\cal E} \right) \\
    + \beta_i \left( \sigma_{\text{a}} - \sigma_{\text{s}} \right) {\cal F}_i \, , \label{eq:CMF_Eddington_approximation_LF_energy_source}
  \end{multline}
  \vspace{-25pt}
  \begin{multline}
    \frac{1}{c^2} \, \partial_t {\cal F}_i + \partial_j {\cal P}_{ij} = - \frac{\sigma_{\text{t}}}{c} {\cal F}_i \\
    + \beta_j \left( \sigma_{\text{t}} \, {\cal P}_{ij} + \sigma_{\text{s}} \, {\cal E} \, \delta_{ij} + \sigma_{\text{a}} \, a_{\text{\tiny R}} \, T^4 \, \delta_{ij} \right) \, , \label{eq:CMF_Eddington_approximation_LF_momentum_source}
  \end{multline}
% \end{linenomath}
  \vspace{-25pt}
% \begin{linenomath}
  \begin{multline}
    {\cal P}_{ij} = \frac{1}{3} \, {\cal E} \, \delta_{ij} \\
    + \frac{1}{c} \left( \beta_j \, {\cal F}_i + \beta_i \, {\cal F}_j - \frac{2}{3} \, \beta_k \, {\cal F}_k \, \delta_{ij} \right) \, . \label{eq:CMF_Eddington_approximation_LF_radiation_pressure}
  \end{multline}
% \end{linenomath}
\end{subequations}
Again, the trace of the radiation pressure returns the Eddington approximation, ${\cal P}_{ii} = {\cal E}$, since the ${\cal O}(\beta)$ term is traceless, as well as symmetric.
These source equations are identical to the LF radiation sources (\ref{eq:lab_frame_source_rate_equations}).
Therefore, the analysis below applies equally well when the Eddington approximation is applied to the LF sources.
Scaling these radiation sources, (\ref{eq:CMF_Eddington_approximation_LF_energy_source}) and (\ref{eq:CMF_Eddington_approximation_LF_momentum_source}), gives:
\begin{subequations}
\label{eq:CMF_Eddington_approximation_radiation_sources}
% \begin{linenomath}
  \begin{multline}
    \partial_t {\cal E} + \frac{1}{\epsilon} \, \partial_i {\cal F}_i = \frac{1}{\epsilon^2} \, \left( \sigma_{\text{t}} - \epsilon \, \sigma_{\text{s}} \right) \, c \, \left( a_{\text{\tiny R}} \, T^4 - {\cal E} \right) \\
    + \frac{1}{\epsilon} \, \beta_i \left( \sigma_{\text{t}} - 2 \, \epsilon \, \sigma_{\text{s}} \right) {\cal F}_i \, , \label{eq:CMF_Eddington_approximation_LF_radiation_energy_source}
  \end{multline}
% \end{linenomath}
  \vspace{-25pt}
% \begin{linenomath}
  \begin{multline}
    \epsilon \, \frac{1}{c^2} \, \partial_t {\cal F}_i + \, \partial_j {\cal P}_{ij} \\
    = - \frac{1}{\epsilon} \, \frac{\sigma_{\text{t}}}{c} \, {\cal F}_i + \beta_j \Bigl[ \sigma_{\text{t}} \, {\cal P}_{ij} + \epsilon \, \sigma_{\text{s}} \, {\cal E} \, \delta_{ij} \Bigr. \\
    + \Bigl. \left( \sigma_{\text{t}} - \epsilon \, \sigma_{\text{s}} \right) a_{\text{\tiny R}} \, T^4 \, \delta_{ij} \Bigr] \, . \label{eq:CMF_Eddington_approximation_LF_radiation_momentum_source} 
  \end{multline}
% \end{linenomath}
\end{subequations}
The radiation-energy source (\ref{eq:CMF_Eddington_approximation_LF_radiation_energy_source}) provides the solution for the scaled radiation energy density,
\begin{subequations}
\label{eqs:LF_grey-Eddington_scaled_vars}
% \begin{linenomath}
  \begin{multline}
    {\cal E} = a_{\text{\tiny R}} \, T^4 \\
    - \epsilon \Biggl[ \frac{\sigma_{\text{s}}}{\sigma_{\text{t}}} \left( a_{\text{\tiny R}} \, T^4 - {\cal E} \right) - \frac{\beta_i}{c} \, {\cal F}_i + \frac{1}{\sigma_{\text{t}} \, c} \, \partial_i {\cal F}_i \Biggr] \, ,
  \end{multline}
% \end{linenomath}
where terms of ${\cal O}(\epsilon^2)$ have been discarded.
The radiation-momentum source (\ref{eq:CMF_Eddington_approximation_LF_radiation_momentum_source}) provides the solution for the scaled radiation flux,
% \begin{linenomath}
  \begin{multline}
    {\cal F}_i = \epsilon \left[ - \frac{c}{\sigma_{\text{t}}} \, \partial_j {\cal P}_{ij} + u_j \, \left( {\cal P}_{ij} + a_{\text{\tiny R}} \, T^4 \, \delta_{ij} \right) \right] \\
    + \epsilon^2 \left[ - \frac{1}{c \, \sigma_{\text{t}}} \, \partial_t {\cal F}_i + u_i \, \frac{\sigma_{\text{s}}}{\sigma_{\text{t}}} \left( {\cal E} - a_{\text{\tiny R}} \, T^4 \right) \right] \, .
  \end{multline}
% \end{linenomath}
% where terms of ${\cal O}(\epsilon^3)$ have been discarded.
Scaling the radiation pressure (\ref{eq:CMF_Eddington_approximation_LF_radiation_pressure}) gives:
% \begin{linenomath}
  \begin{multline}
    {\cal P}_{ij} = \frac{1}{3} \, {\cal E} \, \delta_{ij} \\
    + \epsilon \, \frac{1}{c} \left( \beta_j \, {\cal F}_i + \beta_i \, {\cal F}_j - \frac{2}{3} \, \beta_k \, {\cal F}_k \, \delta_{ij} \right) \, .
  \end{multline}
% \end{linenomath}
\end{subequations}
From these expressions (\ref{eqs:LF_grey-Eddington_scaled_vars}) the zeroth- and first-order solutions are obtained along with the second-order radiation flux:
\begin{subequations}
\label{eqs:CMF_Eddington_EDA_radiation_variables}
% \begin{linenomath}
  \begin{gather}
    {\cal E}^{(0)} = \left[ a_{\text{\tiny R}} \, T^4 \right]^{(0)} \, , \\
    {\cal F}_i^{(0)} = 0 \, , \\
    {\cal P}_{ij}^{(0)} = \left[ \frac{1}{3} \, a_{\text{\tiny R}} \, T^4 \, \delta_{ij} \right]^{(0)} \, , \\
    {\cal E}^{(1)} = \left[ a_{\text{\tiny R}} \, T^4 \right]^{(1)} \, , \\
    {\cal F}_i^{(1)} = \epsilon \left[ - \frac{a_{\text{\tiny R}} \, c}{3 \, \sigma_{\text{t}}} \, \partial_i T^4 + \frac{4}{3} \, u_i \, a_{\text{\tiny R}} \, T^4 \right]^{(0)} \, , \\
    {\cal P}_{ij}^{(1)} = \left[ \frac{1}{3} \, a_{\text{\tiny R}} \, T^4 \, \delta_{ij} \right]^{(1)} \, , \\
    {\cal F}_i^{(2)} = \epsilon \left[ - \frac{a_{\text{\tiny R}} \, c}{3 \, \sigma_{\text{t}}} \, \partial_i T^4 + \frac{4}{3} \, u_i \, a_{\text{\tiny R}} \, T^4 \right]^{(1)} \, .
  \end{gather}
% \end{linenomath}
\end{subequations}
The zeroth- and first-order contributions from the scaled radiation sources (\ref{eq:CMF_Eddington_approximation_radiation_sources}) are:
\begin{subequations}
\label{eqs:CMF_Eddington_approximation_LF_zeroth_first_radiation_sources}
% \begin{linenomath}
\begin{align}
  S_{\textrm{re}}^{(0)} & = \partial_t {\cal E}^{(0)} + \partial_i {\cal F}_i^{(1)} \, , \\
  S_{\textrm{rp}}^{(0)} & = \partial_j {\cal P}_{ij}^{(0)} \, , \\
  S_{\textrm{re}}^{(1)} & = \partial_t {\cal E}^{(1)} + \partial_i {\cal F}_i^{(2)} \, , \\
  S_{\textrm{rp}}^{(1)} & = \frac{1}{c^2} \, \partial_t {\cal F}_i^{(0)} + \partial_j {\cal P}_{ij}^{(1)} \, .
\end{align}
% \end{linenomath}
\end{subequations}
Summing the zeroth- and first-order results for the radiation variables (\ref{eqs:CMF_Eddington_EDA_radiation_variables}) reproduces the EDA (\ref{eq:LF_EDA_radiation_variables}):
\begin{subequations}
% \begin{linenomath}
  \begin{gather}
    {\cal E} = a_{\text{\tiny R}} \, T^4 \, , \\
    {\cal F}_i = - \frac{a_{\text{\tiny R}} \, c}{3 \, \sigma_{\text{t}}} \, \partial_i T^4 + \frac{4}{3} \, u_i \, a_{\text{\tiny R}} \, T^4 \, , \\
    {\cal P}_{ij} = \frac{1}{3} \, a_{\text{\tiny R}} \, T^4 \, \delta_{ij} \, .
  \end{gather}
% \end{linenomath}
\end{subequations}
Summing the zeroth- and first-order contributions from the radiation sources (\ref{eqs:CMF_Eddington_approximation_LF_zeroth_first_radiation_sources}), coupled to the Euler equations (\ref{eqs:RH_equations}), and using the results (\ref{eqs:CMF_Eddington_EDA_radiation_variables}), reproduces the EDA's simplified RH equations (\ref{eqs:EDA_RH_equations}):
\begin{subequations}
% \begin{linenomath}
  \begin{gather}
    \partial_t \rho + \partial_i \left( \rho \, u_i \right) = 0 \, ,
  \end{gather}
  \vspace{-25pt}
  \begin{multline}
    \partial_t \left( \rho \, u_i \right) \\
    + \partial_j \left( \rho \, u_i \, u_j + p_{ij} + \frac{1}{3} \, a_{\text{\tiny R}} \, T^4 \, \delta_{ij} \right) = 0 \, ,
  \end{multline}
  \vspace{-25pt}
  \begin{multline}
    \partial_t \left( E + a_{\text{\tiny R}} \, T^4 \right) \\
    + \partial_i \left[ u_j \left( E_{ij} + p_{ij} + \frac{4}{3} \, a_{\text{\tiny R}} \, T^4 \, \delta_{ij} \right) \right] \\
    = \partial_i \left( \frac{a_{\text{\tiny R}} \, c}{3 \, \sigma_{\text{t}, \text{\tiny R}}} \, \partial_i T^4 \right) \, .
  \end{multline}
% \end{linenomath}
\end{subequations}
Thus, the Eddington approximation preserves the EDA's first-order accuracy.
Further, it does not matter whether the Eddington approximation is applied to the CMF or to the LF radiation sources, the EDA's first-order accuracy is preserved in both situations.
\section{Summary}
\label{sec:summary}
In this work we have derived the EDA from the RH equations via an asymptotic analysis.
Our derivation showed that the EDA and its simplified RH equations are first-order accurate and that transport corrections begin at second order.
Since the EDA is a physical limit of the full set of RH equations it is expected that simplified models of the RH equations should preserve the EDA's first-order accuracy.
We analyzed the grey nonequilibrium-diffusion approximation and the grey Eddington approximation and we showed that they both preserved the EDA's first-order accuracy.
These approximations can be made in the CMF or the LF, and we have shown that the EDA's first-order accuracy is preserved in both cases.
We also presented a test problem in which an equilibrium-diffusion solution was captured from a nonequilibrium-diffusion solver when $\epsilon$ was small.
Other test problems that apply to different RH regimes are needed.
Our results are in agreement with previous asymptotic analyses for neutron transport \cite{HM1975, L1975} and radiative transfer \cite{Morel2000}.
Other analyses \cite{LMM1987, LM1989, LPB1983, MalvagiePomraning1991} for neutron transport and radiative transfer have discussed the effects of initial and boundary conditions, as well as boundary layers, on the asymptotic results.
However, in this paper we have restricted our analysis to the interior solution sufficiently late in time and far away from any boundaries so that their effects on the analysis may be neglected.
An analysis including the initial and boundary conditions, and potentially boundary layers, should be the subject of future work.
Other work should analyze other simplified models and numerical discretizations, and present test problems confirming the analysis, when possible.
It is expected that numerical discretizations which fail to preserve the EDA's first-order accuracy will either fail to produce accurate equilibrium-diffusion solutions or will produce them at a prohibitive computational cost.
A different problem for future work is to investigate multigroup treatments of the RT equation to determine whether they preserve the EDA's first-order accuracy.
\\

\noindent
\textbf{Acknowledgements} One of us (JMF) would like to thank Don Shirk and Bob Singleton for many helpful comments, as well as Scott Doebling for continued support.  This work was performed under the auspices of the US Department of Energy under contract DE-AC52-06NA25396 as LA-UR-17-20878.

%% The Appendices part is started with the command \appendix;
%% appendix sections are then done as normal sections
\appendix
\section{The ${\cal O}(\beta^2)$ LF RT equation}
\label{app:app2C}
This appendix is similar to Section 93 in \cite{MM1999}, where the mixed-frame RT equation with certain CMF functions is presented.
The purpose there and here is to Taylor expand some of the CMF functions so that they depend on the LF frequency instead of the CMF frequency.
However, we retain ${\cal O}(\beta^2)$ terms and scattering terms, both of which are neglected there.
We expand the frequency ratios, the cross sections, and the Planck function through ${\cal O}(\beta^2)$.
While the Planck function and the material cross sections are represented in the CMF, we drop the subscript-o for notational convenience.
We also expand the radiation intensity in the integrand of equation (\ref{app2C:eq:transport}) since it is a function of $\nu^{\, \prime}$, which is a function of $\nu$ by equation (\ref{app2C:eq:primed_and_unprimed}).
Ignoring this expansion produces the wrong results.
We refer to the resulting RT equation as the LF RT equation.
The relativistically exact angle- and frequency-dependent mixed-frame RT equation is:
% \begin{linenomath}
\begin{multline}
\label{app2C:eq:transport}
  \frac{1}{c} \, \partial_t I_{\nu} + \Omega_i \, \partial_i I_{\nu} = - \frac{\nu_{\text{o}}}{\nu} \, \sigma_{\text{t}, \nu_{\text{o}}} I_{\nu} \\
  + \left( \frac{\nu}{\nu_{\text{o}}} \right)^2 \frac{\sigma_{\text{s}}}{4 \pi} \int_{4 \pi} \frac{\nu_{\text{o}}}{\nu^{\, \prime}} \, I_{\nu^{\, \prime}}(\Omega^{\, \prime}) \, d\Omega^{\, \prime} \\
  + \left( \frac{\nu}{\nu_{\text{o}}} \right)^2 \sigma_{\text{a}, \nu_{\text{o}}} \, B_{\nu_{\text{o}}} \, .
\end{multline}
% \end{linenomath}
The ratio of $\nu_{\text{o}}$ to $\nu$ is a function of $\beta$ and $\Omega_i$:
% \begin{linenomath}
\begin{gather}
\label{app2C:eq:nuo_nu_ratio}
  \frac{\nu_{\text{o}}}{\nu} = \gamma_u \left( 1 - \beta_i \, \Omega_i \right) \, ,
\end{gather}
% \end{linenomath}
where the Einstein summation convention is used.
The ratio of LF frequencies is:
% \begin{linenomath}
\begin{gather}
\label{app2C:eq:primed_and_unprimed}
  \frac{\nu^{\, \prime}}{\nu} = \frac{1 - \beta_i \, \Omega_i}{1 - \beta_i \, \Omega_i^{\, \prime}} \, .
\end{gather}
% \end{linenomath}
The Lorentz factor, $\gamma_{\text{u}}$, expanded through ${\cal O}(\beta^2)$ is:
% \begin{linenomath}
\begin{gather}
  \gamma_{\text{u}} = 1 + \frac{1}{2} \, \beta^2 \, ,
\end{gather}
% \end{linenomath}
so the ${\cal O}(\beta^2)$ expansion of the frequency ratios in equation (\ref{app2C:eq:transport}) are:
\begin{subequations}
% \begin{linenomath}
  \begin{gather}
    \frac{\nu_{\text{o}}}{\nu}
    = 1 - \beta_i \, \Omega_i + \frac{1}{2} \, \beta^2 \label{app2C:eq:frequency_ratio} \, , \\
    \frac{\nu_{\text{o}}}{\nu^{\, \prime}}
    = 1 - \beta_i \, \Omega_i^{\, \prime} + \frac{1}{2} \, \beta^2 \, , \\
    \left( \frac{\nu}{\nu_{\text{o}}} \right)^2
    = 1 + 2 \, \beta_i \, \Omega_i + 3 \left( \beta_i \, \Omega_i \right)^2 - \beta^2 \, . \label{app2C:eq:ratios_squared}
  \end{gather}
% \end{linenomath}
\end{subequations}
It is convenient to record here the ${\cal O}(\beta^2)$ expansions of some identities that will be useful when Taylor expanding our functions of interest:
\begin{subequations}
\label{app2C:eq:collected_useful_results}
% \begin{linenomath}
  \begin{gather}
    \nu_{\text{o}} - \nu = \nu \left( - \beta_i \Omega_i + \frac{1}{2} \beta^2 \right) \, , \\
    \left( \nu_{\text{o}} - \nu \right)^2 = \nu^2 \left( \beta_i \Omega_i \right)^2 \, , \\
    \nu = \nu_{\text{o}} \left[ 1 + \beta_i \Omega_i + \left( \beta_i \Omega_i \right)^2 - \frac{1}{2} \beta^2 \right] \, , \\
    \partial_{\nu_{\text{o}}} \nu = 1 + \beta_i \Omega_i + \left( \beta_i \Omega_i \right)^2 - \frac{1}{2} \beta^2 = \frac{\nu}{\nu_{\text{o}}} \, , \\
    \frac{\nu \left( \nu_{\text{o}} - \nu \right) }{\nu_{\text{o}}} = \nu \left[ - \beta_i \Omega_i - \left( \beta_i \Omega_i \right)^2 + \frac{1}{2} \beta^2 \right] \, , \\
    \left( \nu_{\text{o}} - \nu \right)^2 f(\nu_{\text{o}}) = \left( \nu_{\text{o}} - \nu \right)^2 f(\nu) \, .
  \end{gather}
% \end{linenomath}
\end{subequations}
The Taylor-expansion of a general function of the CMF frequency, with respect to the LF frequency, through ${\cal O}((\nu_{\text{o}} - \nu)^2) \sim {\cal O}(\beta^2)$, is:
% \begin{linenomath}
\begin{multline}
\label{app2C:eq:Taylor-expansion}
  f(\nu_{\text{o}})
  = f + \left( \nu_{\text{o}} - \nu \right) \partial_{\nu_{\text{o}}} f + \frac{1}{2} \left( \nu_{\text{o}} - \nu \right)^2 \partial_{\nu_{\text{o}}}^2 f \\
  = f - \beta_i \Omega_i \nu \partial_{\nu} f + \frac{1}{2} \beta_i \beta_j \left( \left( \delta_{ij} - 2 \Omega_i \Omega_j \right) \nu \partial_{\nu} f \right. \\
  + \left. \Omega_i \Omega_j \nu^2 \partial_{\nu}^2 f \right) \, .
\end{multline}
% \end{linenomath}
The ${\cal O}(\beta^2)$ Taylor expansions of the total cross-section and the Planck function, and the product of the absorption cross-section with the Planck function, are:
\begin{subequations}
\label{eqs:Taylor_expansions}
% \begin{linenomath}
  \begin{multline}
    \sigma_{\text{t}, \nu_{\text{o}}} = \sigma_{\text{t}, \nu} - \beta_i \Omega_i \nu \partial_{\nu} \sigma_{\text{t}, \nu} \\
    + \frac{1}{2} \beta_i \beta_j \left( \left( \delta_{ij} - 2 \Omega_i \Omega_j \right) \nu \partial_{\nu} \sigma_{\text{t}, \nu} \right. \\
    + \left. \Omega_i \Omega_j \nu^2 \partial_{\nu}^2 \sigma_{\text{t}, \nu} \right) \, ,
  \end{multline}
  \vspace{-25pt}
  \begin{multline}
    B_{\nu_{\text{o}}} = B_{\nu} - \beta_i \Omega_i \nu \partial_{\nu} B_{\nu} \\
    + \frac{1}{2} \beta_i \beta_j \left( \left( \delta_{ij} - 2 \Omega_i \Omega_j \right) \nu \partial_{\nu} B_{\nu} \right. \\
    + \left. \Omega_i \Omega_j \nu^2 \partial_{\nu}^2 B_{\nu} \right) \, , \label{app:eq:Planck_LF_nu} 
  \end{multline}
  \vspace{-25pt}
  \begin{multline}
    \sigma_{\text{a}, \nu_{\text{o}}} B_{\nu_{\text{o}}} = \sigma_{\text{a}, \nu} B_{\nu} - \beta_i \Omega_i \nu \partial_{\nu} \left( \sigma_{\text{a}, \nu} B_{\nu} \right) \\
    + \frac{1}{2} \beta_i \beta_j \left( \left( \delta_{ij} - 2 \Omega_i \Omega_j \right) \nu \partial_{\nu} \left( \sigma_{\text{a}, \nu} B_{\nu} \right) \right. \\
    + \left. \Omega_i \Omega_j \nu^2 \partial_{\nu}^2 \left( \sigma_{\text{a}, \nu} B_{\nu} \right) \right) \, .
  \end{multline}
% \end{linenomath}
\end{subequations}
The Taylor expansion of the LF radiation intensity, in the integrand of equation (\ref{app2C:eq:transport}), proceeds along the same lines.
However, the CMF frequency in the previous expressions is now a LF frequency, $\nu^{\, \prime}$, and the necessary relations take a slightly different form through ${\cal O}(\beta^2)$:
\begin{subequations}
\label{app2C:eq:new_relations}
% \begin{linenomath}
  \begin{gather}
    \nu = \nu^{\, \prime} \left( 1 + \beta_i \left( \Omega_i^{\, \prime} - \Omega_i \right) \right) \, , \\
    \left( \nu^{\, \prime} - \nu \right) = \nu \beta_i \left( \Omega_i^{\, \prime} - \Omega_i \right) \, , \\
    \left( \nu^{\, \prime} - \nu \right)^2 = \nu^2 \beta_i \beta_j \left( \Omega_i^{\, \prime} - \Omega_i \right) \left( \Omega_j^{\, \prime} - \Omega_j \right) \, , \\
    \partial_{\nu^{\, \prime}} \nu = 1 + \beta_i \left( \Omega_i^{\, \prime} - \Omega_i \right) = \frac{\nu}{\nu^{\, \prime}} \, , \\
    \left( \nu^{\, \prime} - \nu \right) \frac{\nu}{\nu^{\, \prime}} = \nu \beta_i \left( \Omega_i^{\, \prime} - \Omega_i \right) \, .
  \end{gather}
% \end{linenomath}
\end{subequations}
In arriving at the relations above, we have used the fact that $\beta_i \beta_j \Omega_i (\Omega_j - \Omega_j^{\, \prime})$ is zero at ${\cal O}(\beta^2)$ since the angular variables are then the same.
The ${\cal O}(\beta^2)$ Taylor expansion is:
% \begin{linenomath}
\begin{multline}
  f(\nu^{\, \prime})
\nonumber
  = f + \left( \nu^{\, \prime} - \nu \right) \left( \frac{\nu}{\nu^{\, \prime}} \right) \partial_{\nu} f \\
  + \frac{1}{2} \left( \nu^{\, \prime} - \nu \right)^2 \left( \frac{1}{\nu} \partial_{\nu} f + \partial_{\nu}^2 f \right) \\
  = f + \beta_i \left( \Omega_i^{\, \prime} - \Omega_i \right) \nu \partial_{\nu} f \, ,
\end{multline}
% \end{linenomath}
such that the Taylor expanded radiation intensity, through ${\cal O}(\beta^2)$, is
% \begin{linenomath}
\begin{gather}
\label{eq:Taylor_expand_I}
  I_{\nu}(\Omega^{\, \prime}) = I_{\nu} + \beta_i \left( \Omega_i^{\, \prime} - \Omega_i \right) \nu \partial_{\nu} I_{\nu} \, .
\end{gather}
% \end{linenomath}
% Now that all of the Taylor-expansions have been performed it is necessary to combine them appropriately according to the three terms on the RHS of equation (\ref{app2C:eq:transport}).
We now combine the results in equations (\ref{eqs:Taylor_expansions}) and (\ref{eq:Taylor_expand_I}) to use in the three terms on the RHS of equation (\ref{app2C:eq:transport}).
The first term is straight-forward:
% \begin{linenomath}
\begin{multline}
  - \frac{\nu_{\text{o}}}{\nu} \, \sigma_{\text{t}, \nu_{\text{o}}} \, I_{\nu}
  = - \sigma_{\text{t}, \nu} \, I_{\nu} + \beta_i \, \Omega_i \left( \sigma_{\text{t}, \nu} \, I_{\nu} + I_{\nu} \, \nu \, \partial_{\nu} \sigma_{\text{t}, \nu} \right) \\
  - \frac{1}{2} \, \beta_i \, \beta_j \left( \sigma_{\text{t}, \nu} \, I_{\nu} \, \delta_{ij} + I_{\nu} \, \nu \, \partial_{\nu} \sigma_{\text{t}, \nu} \, \delta_{ij} \right. \\
  + \left. \Omega_i \, \Omega_j \, I_{\nu} \, \nu^2 \, \partial_{\nu}^2 \sigma_{\text{t}, \nu} \right) \, . \label{app2C:eq:first_term}
\end{multline}
% \end{linenomath}
The second term is best analyzed by breaking it into parts.
The integrand is
% \begin{linenomath}
\begin{multline}
  \frac{\nu_{\text{o}}}{\nu^{\, \prime}} \, I_{\nu^{\, \prime}}(\Omega^{\, \prime}) = I^{\, \prime}_{\nu} \\
  + \beta_i \left( - \Omega_i^{\, \prime} \, I^{\, \prime}_{\nu} + \left( \Omega_i^{\, \prime} - \Omega_i \right) \nu \, \partial_{\nu} I^{\, \prime}_{\nu} \right) \\
  + \frac{1}{2} \, \beta^2 \, I^{\, \prime}_{\nu} \, ,
\end{multline}
% \end{linenomath}
where we have written $I^{\, \prime}_{\nu} = I_{\nu}(\Omega^{\, \prime})$ for notational convenience.
The result of the angular integral is
% \begin{linenomath}
\begin{multline}
  \int_{4 \pi} \frac{\nu_{\text{o}}}{\nu^{\, \prime}} \, I_{\nu^{\, \prime}}(\Omega^{\, \prime}) \, d\Omega^{\, \prime} = \phi_{\nu} \\
  + \beta_i \left( - F_{\nu, i} + \nu \, \partial_{\nu} F_{\nu, i} - \Omega_i \, \nu \, \partial_{\nu} \phi_{\nu} \right) \\
  + \frac{1}{2} \, \beta^2 \phi_{\nu} \, .
\end{multline}
% \end{linenomath}
The ratio $\left( \nu / \nu_{\text{o}} \right)^2$ multiplying the integral is given in equation (\ref{app2C:eq:ratios_squared}), and the second term of equation (\ref{app2C:eq:transport}) is now written as:
% \begin{linenomath}
\begin{multline}
  \left( \frac{\nu}{\nu_{\text{o}}} \right)^2 \frac{\sigma_{\text{s}}}{4 \pi} \int_{4 \pi} \frac{\nu_{\text{o}}}{\nu^{\, \prime}} \, I_{\nu^{\, \prime}}(\Omega^{\, \prime}) \, d\Omega^{\, \prime} = \frac{\sigma_{\text{s}}}{4 \pi} \, \phi_{\nu} \\
  + \frac{\sigma_{\text{s}}}{4 \pi} \, \beta_i \left( 2 \, \Omega_i \, \phi_{\nu} - \Omega_i \, \nu \, \partial_{\nu} \phi_{\nu} - F_{\nu, i} + \nu \, \partial_{\nu} F_{\nu, i} \right) \\
  + \frac{\sigma_{\text{s}}}{4 \pi} \, \beta_i \, \beta_j \left[ \left( 3 \, \Omega_i \, \Omega_j - \frac{1}{2} \, \delta_{ij} \right) \phi_{\nu} - 2 \, \Omega_i \, \Omega_j \, \nu \, \partial_{\nu} \phi_{\nu} \right] \\
  - \frac{2 \, \sigma_{\text{s}}}{4 \pi} \, \beta_i \, \beta_j \, \Omega_i \left( F_{\nu, j} - \nu \, \partial_{\nu} F_{\nu, j} \right) \, . \label{app2C:eq:second_term}
\end{multline}
% \end{linenomath}
The third term is:
% \begin{linenomath}
\begin{multline}
  \left( \frac{\nu}{\nu_{\text{o}}} \right)^2 \sigma_{\text{a}, \nu_{\text{o}}} \, B_{\nu_{\text{o}}}(T) = \sigma_{\text{a}, \nu} \, B_{\nu} \hfill \\
  \hfill + \beta_i \, \Omega_i \left[ 2 \, \sigma_{\text{a}, \nu} \, B_{\nu} - \nu \, \partial_{\nu} \left( \sigma_{\text{a}, \nu} \, B_{\nu} \right) \right] \\
  + \beta_i \, \beta_j \Biggl[ \left( 3 \, \Omega_i \, \Omega_j - \delta_{ij} \right) \sigma_{\text{a}, \nu} \, B_{\nu} + \frac{1}{2} \, \Omega_i \, \Omega_j \, \nu^2 \, \partial_{\nu}^2 \left( \sigma_{\text{a}, \nu} \, B_{\nu} \right) \Biggr. \\
  \hfill + \Biggl. \left( \frac{1}{2} \, \delta_{ij} - 3 \, \Omega_i \, \Omega_j \right) \nu \, \partial_{\nu} \left( \sigma_{\text{a}, \nu} \, B_{\nu} \right) \Biggr] \, . \label{app2C:eq:third_term}
\end{multline}
% \end{linenomath}
Collecting equations (\ref{app2C:eq:first_term}), (\ref{app2C:eq:second_term}) and (\ref{app2C:eq:third_term}), the ${\cal O}(\beta^2)$ LF RT equation is:
% \begin{linenomath}
\begin{multline}
  \frac{1}{c} \partial_t I_{\nu} + \Omega_i \partial_i I_{\nu} = \frac{\sigma_{\text{s}}}{4 \pi} \phi_{\nu} + \sigma_{\text{a}, \nu} B_{\nu} - \sigma_{\text{t}, \nu} I_{\nu} \hfill \\
  + \beta_i \Omega_i \left[ \sigma_{\text{t}, \nu} I_{\nu} + I_{\nu} \nu \partial_{\nu} \sigma_{\text{t}, \nu} + 2 \frac{\sigma_{\text{s}}}{4 \pi} \phi_{\nu} + 2 \sigma_{\text{a}, \nu} B_{\nu} \right. \hfill \\
  - \left. \frac{\sigma_{\text{s}}}{4 \pi} \nu \partial_{\nu} \phi_{\nu} - \nu \partial_{\nu} \left( \sigma_{\text{a}, \nu} B_{\nu} \right) \right] - \frac{\sigma_{\text{s}}}{4 \pi} \beta_i \left( F_{\nu, i} - \nu \partial_{\nu} F_{\nu, i} \right) \\
  - \frac{1}{2} \beta_i \beta_j \left( \sigma_{\text{t}, \nu} \delta_{ij} I_{\nu} + \delta_{ij} I_{\nu} \nu \partial_{\nu} \sigma_{\text{t}, \nu} + \Omega_i \Omega_j I_{\nu} \nu^2 \partial_{\nu}^2 \sigma_{\text{t}, \nu} \right) \\
  + \frac{\sigma_{\text{s}}}{4 \pi} \beta_i \beta_j \Biggl[ \left( 3 \Omega_i \Omega_j - \frac{1}{2} \delta_{ij} \right) \phi_{\nu} - 2 \Omega_i \Omega_j \nu \partial_{\nu} \phi_{\nu} \Biggr. \hfill \\
  \hfill - \Biggl. 2 \Omega_i F_{\nu, j} + 2 \Omega_i \nu \partial_{\nu} F_{\nu, j} \Biggr] \\
  + \beta_i \beta_j \Biggl[ \left( 3 \Omega_i \Omega_j - \delta_{ij} \right) \sigma_{\text{a}, \nu} B_{\nu} + \frac{1}{2} \Omega_i \Omega_j \nu^2 \partial_{\nu}^2 \left( \sigma_{\text{a}, \nu} B_{\nu} \right) \Biggr. \\ 
  \hfill + \Biggl. \left( \frac{1}{2} \delta_{ij} - 3 \Omega_i \Omega_j \right) \nu \partial_{\nu} \left( \sigma_{\text{a}, \nu} B_{\nu} \right) \Biggr] \, . \label{app:eq:the_goal}
\end{multline}
% \end{linenomath}
This result may be compared with equation 93.4 of \cite{MM1999}, although there the ${\cal O}(\beta^2)$ terms are neglected as are the scattering cross sections.
We believe this is the first time this equation has been presented in the literature.
\section{The ${\cal O}(\epsilon^2)$ analysis}
\label{app:app3A}
In this appendix we scale and analyze the ${\cal O}(\beta^2)$ LF RT equation (\ref{app:eq:the_goal}).
First, we write $\sigma_{\text{a}, \nu} = \sigma_{\text{t}, \nu} - \sigma_{\text{s}}$, and we reiterate that $\sigma_{\text{s}}$ is frequency-independent.
% The asymptotic analysis of the RT equation (\ref{app:eq:the_result}) requires implementing the scaled ratios (\ref{eq:scalings}) and retaining terms through ${\cal O}(\epsilon^2)$.
The scaled LF RT equation is:
\begin{subequations}
\label{app3a:eq:scaled_RT_equation_and_solution}
% \begin{linenomath}
\begin{multline}
\label{app3A:eq:scaled_RT}
  \epsilon^2 \frac{1}{c} \, \partial_t I_{\nu} + \epsilon \, \Omega_i \, \partial_i I_{\nu} = \sigma_{\text{t}, \nu} \left( B_{\nu} - I_{\nu} \right) \\
  + \epsilon \, \biggl\{ \frac{\sigma_{\text{s}}}{4 \pi} \left( \phi_{\nu} - 4 \pi \, B_{\nu} \right) + \beta_i \, \Omega_i \Bigl[ \sigma_{\text{t}, \nu} \, I_{\nu} + I_{\nu} \, \nu \, \partial_{\nu} \sigma_{\text{t}, \nu} \Bigr. \biggr. \hfill \\
  + \biggl. \Bigl. 2 \, \sigma_{\text{t}, \nu} \, B_{\nu} - \nu \, \partial_{\nu} \left( \sigma_{\text{t}, \nu} \, B_{\nu} \right) \Bigr] \biggr\} \, + \, \epsilon^2 \, \Biggl\{ \frac{\sigma_{\text{s}}}{4 \pi} \, \beta_i \Bigl[ - F_{\nu, i} + \nu \, \partial_{\nu} F_{\nu, i} \Bigr. \Biggr. \\
  \hfill + \Biggl. \Bigl. \Omega_i \left( 2 \, \phi_{\nu} - 8 \pi \, B_{\nu} - \nu \, \partial_{\nu} \phi_{\nu} + 4 \pi \, \nu \, \partial_{\nu} B_{\nu} \right) \Bigr] \Biggr. \\
  + \Biggr. \beta_i \, \beta_j \Biggl[ \left( 3 \, \Omega_i \, \Omega_j - \delta_{ij} \right) \sigma_{\text{t}, \nu} \, B_{\nu} + \frac{1}{2} \, \Omega_i \, \Omega_j \, \nu^2 \, \partial_{\nu}^2 \left( \sigma_{\text{t}, \nu} \, B_{\nu} \right) \Biggr. \Biggr. \hfill \\
  + \Biggl. \Biggl. \left( \frac{1}{2} \, \delta_{ij} - 3 \, \Omega_i \, \Omega_j \right) \nu \, \partial_{\nu} \left( \sigma_{\text{t}, \nu} \, B_{\nu} \right) - \frac{1}{2} \, \Bigl( \sigma_{\text{t}, \nu} \, I_{\nu} \, \delta_{ij} \Bigr. \Biggr. \Biggr. \hfill \\
  \hfill + \Biggl. \Biggl. \Bigl. \delta_{ij} \, I_{\nu} \, \nu \, \partial_{\nu} \sigma_{\text{t}, \nu} + \Omega_i \, \Omega_j \, I_{\nu} \, \nu^2 \, \partial_{\nu}^2 \sigma_{\text{t}, \nu} \Bigr) \Biggr] \Biggr\} \, .
\end{multline}
% \end{linenomath}
This can be rearranged to produce the ${\cal O}(\epsilon^2)$ solution for the radiation intensity:
% \begin{linenomath}
\begin{multline}
\label{app3A:eq:scaled_RT_solution}
  I_{\nu} = B_{\nu} + \epsilon \left\{ - \frac{1}{\sigma_{\text{t}, \nu}} \, \Omega_i \, \partial_i I_{\nu} + \beta_i \, \Omega_i \Bigl( I_{\nu} + 2 \, B_{\nu} \Bigr. \right. \hfill \\
  \hfill \left. \Bigl. + \frac{\nu}{\sigma_{\text{t}, \nu}} \bigl[ I_{\nu} \, \partial_{\nu} \sigma_{\text{t}, \nu} - \partial_{\nu} \left( \sigma_{\text{t}, \nu} \, B_{\nu} \right) \bigr] \Bigr) + \frac{\sigma_{\text{s}}}{\sigma_{\text{t}, \nu}} \left( \frac{\phi_{\nu}}{4 \pi} - B_{\nu} \right) \right\} \\
  + \epsilon^2 \left\{ - \frac{1}{c \, \sigma_{\text{t}, \nu}} \, \partial_t I_{\nu} - \frac{\sigma_{\text{s}}}{\sigma_{\text{t}, \nu}} \beta_i \left[ \frac{1}{4 \pi} \left( F_{\nu, i} - \nu \, \partial_{\nu} F_{\nu, i} \right) \right. \right. \hfill \\
  \hfill - \left. \left. \Omega_i \left( \frac{2 \, \phi_{\nu}}{4 \pi} - \frac{\nu}{4 \pi} \, \partial_{\nu} \phi_{\nu} - 2 \, B_{\nu} + \nu \, \partial_{\nu} B_{\nu} \right) \right] \right. \\
  + \left. \beta_i \, \beta_j \left[ \left( 3 \, \Omega_i \, \Omega_j - \delta_{ij} \right) B_{\nu} - \frac{I_{\nu}}{2} \left( 1 + \frac{\nu}{\sigma_{\text{t}, \nu}} \, \partial_{\nu} \sigma_{\text{t}, \nu} \right) \delta_{ij} \right. \right. \\
  + \left. \left. \frac{1}{2 \, \sigma_{\text{t}, \nu}} \left( \delta_{ij} - 6 \, \Omega_i \, \Omega_j \right) \nu \, \partial_{\nu} \left( \sigma_{\text{t}, \nu} \, B_{\nu} \right) \right. \right. \\
  \left. \left. + \frac{\Omega_i \, \Omega_j \, \nu^2}{2 \, \sigma_{\text{t}, \nu}} \left[ \partial_{\nu}^2 \left( \sigma_{\text{t}, \nu} \, B_{\nu} \right) - I_{\nu} \, \partial_{\nu}^2 \sigma_{\text{t}, \nu} \right] \right] \right\} \, .
\end{multline}
% \end{linenomath}
\end{subequations}
% It is worth noticing a few things about equation (\ref{app3A:eq:scaled_RT_solution}) which will affect the order of angle- and frequency-integrations to be made.
% First, the Planck function $B_{\nu}$, and the angle-integrated radiation variables $\phi_{\nu}$ and $F_{\nu, i}$, are angle independent such that odd-angular moments of these variables are identically zero.
% Second, there exist a few terms where the frequency derivative is multiplied by frequency, and integration-by-parts in the frequency variable will help in the analysis of most of these terms.
% As for the integration-by-parts in the frequency variable, it is assumed that all radiation variables have compact support over the frequency domain $\nu \in [0, \infty)$, such that the radiation variables are zero at the frequency boundaries:
% \begin{subequations}
% % \begin{linenomath}
%   \begin{gather}
%     I_{\nu = 0} = 0 = I_{\nu = \infty} \, , \\
%     B_{\nu = 0} = 0 = B_{\nu = \infty} \, , \\
%     \phi_{\nu = 0} = 0 = \phi_{\nu = \infty} \, , \\
%     F_{\nu = 0, i} = 0 = F_{\nu = \infty, i} \, .
%   \end{gather}
% % \end{linenomath}
% \end{subequations}
% Based on these comments, we will first perform angle integrations and then frequency integrations of the radiation intensity when constructing its angle- and frequency-integrated angular moments.

The zeroth-, first- and second-order solutions of the radiation intensity can now be determined.
Their frequency-integrated angular moments produce the associated radiation variables.
The zeroth-order solutions are:
% $\quad$ \\
% \underline{${\cal O}(0)$} \\
% The leading-order RT equation and its solution, from equations (\ref{app3a:eq:scaled_RT_equation_and_solution}), are
% \begin{subequations}
% % \begin{linenomath}
%   \begin{gather}
%     \left[ \sigma_{\text{t}} \left( B_{\nu} - I_{\nu} \right) \right]^{(0)} = 0 \, , \\
%     I_{\nu}^{(0)} = B_{\nu}^{(0)} \, ,
%   \end{gather}
% % \end{linenomath}
% \end{subequations}
% such that the zeroth-order angle-integrated, and angle- and frequency-integrated, angular moments are:
\begin{subequations}
\label{app3A:eq:leading_order_results}
% \begin{linenomath}
  \begin{gather}
    I_{\nu}^{(0)} = B_{\nu}^{(0)} \, , \\
    I^{(0)} = \int_0^{\infty} I_{\nu}^{(0)} d\nu = \left[ \frac{a_{\text{\tiny R}} \, c}{4 \pi} \, T^4 \right]^{(0)} \, , \\
%     \phi_{\nu}^{(0)} = \int_{4 \pi} I_{\nu}^{(0)} \, d\Omega = \left[ 4 \pi \, B_{\nu} \right]^{(0)} \, , \\
%     F_{\nu, i}^{(0)} = \int_{4 \pi} \Omega_i \, I_{\nu}^{(0)} \, d\Omega = 0 \, , \\
%     P_{\nu, ij}^{(0)} = \int_{4 \pi} \Omega_i \, \Omega_j \, I_{\nu}^{(0)} \, d\Omega = \left[ \frac{4 \pi}{3} \, B_{\nu} \, \delta_{ij} \right]^{(0)} \, , \\% = \frac{1}{3} \delta_{ij} \phi_{\nu}^{(0)} \, , \\
    {\cal E}^{(0)} = \frac{1}{c} \int_{4 \pi} I^{(0)} d\Omega = \left[ a_{\text{\tiny R}} \, T^4 \right]^{(0)} \, , \\
    {\cal F}_i^{(0)} = \int_{4 \pi} \Omega_i \, I^{(0)} d\Omega = 0 \, , \\
    {\cal P}_{ij}^{(0)} = \frac{1}{c} \int_{4 \pi} \Omega_i \, \Omega_j \, I^{(0)} d\Omega = \left[ \frac{1}{3} \, {\cal E} \, \delta_{ij} \right]^{(0)} \, .
%     {\cal P}_{ij}^{(0)} = \frac{1}{c} \int_{4 \pi} \Omega_i \, \Omega_j \, I^{(0)} d\Omega = \left[ \frac{a_{\text{\tiny R}} \, T^4}{3} \, \delta_{ij} \right]^{(0)} \, .
  \end{gather}
% \end{linenomath}
\end{subequations}
% $\quad$ \\
% \underline{${\cal O}(1)$} \\
The first-order solutions using these results are:
\begin{subequations}
\label{app3A:eq:first_order_results}
% \begin{linenomath}
\begin{multline}
%   \left[ \sigma_{\text{t}} \left( B_{\nu} - I_{\nu} \right) \right]^{(1)} + \Omega_i \left[ - \Omega_i \partial_i B_{\nu} + \beta_i \left( 3 \sigma_{\text{t}} B_{\nu} - \sigma_{\text{t}} \nu \partial_{\nu} B_{\nu} \right) \right]^{(0)} = 0 \, , \\
  I_{\nu}^{(1)} = B_{\nu}^{(1)} + \Omega_i \Bigl[ - \frac{1}{\sigma_{\text{t}}} \, \partial_i B_{\nu} \Bigr. \hfill \\
  \hfill + \Bigl. \beta_i \left( 3 \, B_{\nu} - \nu \, \partial_{\nu} B_{\nu} \right) \Bigr]^{(0)} \, , \label{app3A:eq:nu_dept_I} 
\end{multline}
% % \end{linenomath}
% \end{subequations}
% such that the first-order angle-integrated, and angle- and frequency-integrated, angular moments are:
% \begin{subequations}
% % \begin{linenomath}
\vspace{-25pt}
\begin{multline}
  I^{(1)} = \int_0^{\infty} I_{\nu}^{(1)} d\nu = \left[ \frac{a_{\text{\tiny R}} \, c}{4 \pi} \, T^4 \right]^{(1)} \hfill \\
  \hfill + \frac{\Omega_i}{4 \pi} \left[ - \frac{a_{\text{\tiny R}} \, c}{\sigma_{\text{t}, \text{\tiny R}}} \, \partial_i T^4 + 4 \, u_i \, a_{\text{\tiny R}} \, T^4 \right]^{(0)} \, , \label{app3A:eq:I}
\end{multline}
%   \phi_{\nu}^{(1)} = \int_{4 \pi} I_{\nu}^{(1)} \, d\Omega = 4 \pi B_{\nu}^{(1)} \, , \\
%   F_{\nu, i}^{(1)} = \int_{4 \pi} \Omega_i I_{\nu}^{(1)} \, d\Omega = \frac{4 \pi}{3} \left[ - \frac{1}{\sigma_{\text{t}}} \partial_i B_{\nu} + \beta_i \left( 3 B_{\nu} - \nu \partial_{\nu} B_{\nu} \right) \right]^{(0)} \, , \label{app3A:eq:first_order_F} \\
%   P_{\nu, ij}^{(1)} = \int_{4 \pi} \Omega_i \Omega_j I_{\nu}^{(1)} \, d\Omega = \frac{4 \pi}{3} \delta_{ij} B_{\nu}^{(1)} = \frac{1}{3} \delta_{ij} \phi_{\nu}^{(1)} \, , \label{app3A:eq:first_order_P} \\
\vspace{-15pt}
\begin{gather}
  {\cal E}^{(1)} = \frac{1}{c} \int_{4 \pi} I^{(1)} d\Omega = \left[ a_{\text{\tiny R}} \, T^4 \right]^{(1)} \, ,
\end{gather}
\vspace{-15pt}
\begin{align}
\nonumber
  {\cal F}_i^{(1)}
  & = \int_{4 \pi} \Omega_i \, I^{(1)} d\Omega \\
  & = \left[ - \frac{a_{\text{\tiny R}} \, c}{3 \, \sigma_{\text{t}, \text{\tiny R}}} \, \partial_i T^4 + \frac{4}{3} \, a_{\text{\tiny R}} \, u_i T^4 \right]^{(0)} \, , \label{app3A:eq:first_order_calF}
\end{align}
\vspace{-15pt}
\begin{gather}
  {\cal P}_{ij}^{(1)} = \frac{1}{c} \int_{4 \pi} \Omega_i \, \Omega_j \, I^{(1)} d\Omega = \left[ \frac{1}{3} \, {\cal E} \, \delta_{ij} \right]^{(1)} \, . \label{app3A:eq:first_order_calP}
\end{gather}
% \end{linenomath}
\end{subequations}
% In passing from equation (\ref{app3A:eq:nu_dept_I}) to (\ref{app3A:eq:I}) we have used the definition of the Rosseland-averaged cross section, $\sigma_{\text{t}, \text{\tiny R}}$.
% The definition is in \cite{MM1999}.%, has been defined,
The Rosseland-averaged cross section \cite{MM1999}, $\sigma_{\text{t}, \text{\tiny R}}$, has been used in passing from equation (\ref{app3A:eq:nu_dept_I}) to (\ref{app3A:eq:I}).
% % \begin{linenomath}
% \begin{align}
% \nonumber
%   \int_0^{\infty} \frac{1}{\sigma_{\text{t}}} \partial_i B_{\nu} \, d\nu
%   & = \partial_i T \int_0^{\infty} \frac{1}{\sigma_{\text{t}}} \partial_T B_{\nu} \, d\nu 
%   = \partial_i T \underbrace{\frac{\int_0^{\infty} \frac{1}{\sigma_{\text{t}}} \partial_T B_{\nu}}{\int_0^{\infty} \partial_T B_{\nu} \, d\nu}}_{\equiv \frac{1}{\sigma_{\text{t}, \text{\tiny R}}}} \underbrace{\int_0^{\infty} \partial_T B_{\nu} \, d\nu}_{= \partial_T \int_0^{\infty} B_{\nu} \, d\nu} \\
% \nonumber
%   & = \frac{1}{\sigma_{\text{t}, \text{\tiny R}}} \partial_i T \left( \partial_T \frac{a_{\text{\tiny R}} c T^4}{4 \pi} \right) = \frac{a_{\text{\tiny R}} c}{4 \pi \sigma_{\text{t}, \text{\tiny R}}} \partial_i T^4 \, ,
% \end{align}
% % \end{linenomath}
% and an integration-by-parts has been performed in the frequency variable.
% $\quad$ \\
% \underline{$O(2)$} \\
% The second-order RT equation and its solution, from equations (\ref{app3a:eq:scaled_RT_equation_and_solution}), simplified using the results presented above in (\ref{app3A:eq:leading_order_results}) and (\ref{app3A:eq:first_order_results}), are
The second-order radiation-intensity solution, using the results in (\ref{app3A:eq:leading_order_results}) and (\ref{app3A:eq:first_order_results}), is:
\begin{subequations}
% \begin{linenomath}
\begin{multline}
\label{app3A:eq:second_order_I}
%     & \left[ \sigma_{\text{t}} \left( B_{\nu} - I_{\nu} \right) \right]^{(2)} + \Omega_i \left[ - \partial_i B_{\nu} + \beta_i \sigma_{\text{t}} \left( 3 B_{\nu} - \nu \partial_{\nu} B_{\nu} \right) \right]^{(1)} = - \frac{1}{c} \partial_t B_{\nu}^{(0)} \\
%     & + \Omega_i \Omega_j \left[ - \partial_i \left( \frac{1}{\sigma_{\text{t}}} \partial_j B_{\nu} \right) + \beta_i \partial_j B_{\nu} + \frac{1}{\sigma_{\text{t}}} \nu \left( \partial_{\nu} \sigma_{\text{t}} \right) \beta_i \partial_j B_{\nu} + \partial_i \left( \beta_j \left( 3 B_{\nu} - \nu \partial_{\nu} B_{\nu} \right) \right) \right]^{(0)} \\
%     & + \left[ \frac{1}{2} \beta^2 \sigma_{\text{t}} \left( 3 B_{\nu} - \nu \partial_{\nu} B_{\nu} \right) - \beta_i \beta_j \Omega_i \Omega_j \sigma_{\text{t}} \left( 6 B_{\nu} - 4 \nu \partial_{\nu} B_{\nu} + \frac{1}{2} \nu^2 \partial_{\nu}^2 B_{\nu} \right) \right]^{(0)} \, , \\
  I_{\nu}^{(2)} = B_{\nu}^{(2)} \, + \, \Omega_i \left\{ - \frac{1}{\sigma_{\text{t}}} \, \partial_i B_{\nu} + \beta_i \left( 3  \,B_{\nu} - \nu  \,\partial_{\nu} B_{\nu} \right) \right\}^{(1)} \hfill \\
  + \left\{ - \frac{1}{c \, \sigma_{\text{t}}} \, \partial_t B_{\nu} - \frac{1}{2} \, \beta^2 \left( 3 \, B_{\nu} - \nu \, \partial_{\nu} B_{\nu} \right) \right. \hfill \\
  \hfill + \left. \Omega_i \Omega_j \left[ \beta_j \left( - \frac{1}{\sigma_{\text{t}}} \left( 1 + \frac{1}{\sigma_{\text{t}}} \, \nu \, \partial_{\nu} \sigma_{\text{t}} \right) \partial_i B_{\nu} \right. \right. \right. \\
  + \left. \left. \left. \beta_i \left( 6 \, B_{\nu} - 4 \, \nu \, \partial_{\nu} B_{\nu} + \frac{1}{2} \, \nu^2 \, \partial_{\nu}^2 B_{\nu} \right) \right) \right] \right\}^{(0)} \, . \hfill
\end{multline}
% \end{linenomath}
\end{subequations}
This solution cannot be frequency integrated since the functional form of the total cross section is unknown.
Previously, integration by parts was used to move the frequency-derivatives, but that does not work now.
However, the frequency-dependent radiation flux can be constructed by taking the first angular moment of (\ref{app3A:eq:second_order_I}):
% The second-order radiation flux is:
\begin{subequations}
% \begin{linenomath}
\begin{multline}
  {\cal F}_{\nu, i} = \int_{4 \pi} \Omega_i \, I_{\nu}^{(2)} d\Omega = \frac{4 \pi}{3} \biggl[ - \frac{1}{\sigma_{\text{t}}} \, \partial_i B_{\nu} \biggr. \hfill \\
  \hfill + \biggl. \beta_i \left( 3 \, B_{\nu} - \nu \, \partial_{\nu} B_{\nu} \right) \biggr]^{(1)} \, .
\end{multline}
This expression can be frequency-integrated to produce the second-order radiation flux:
\begin{multline}
%   F_{\nu, i}^{(2)}
%   & = \frac{4 \pi}{3} \left[ - \frac{1}{\sigma_{\text{t}}} \partial_i B_{\nu} + \beta_i \left( 3 B_{\nu} - \nu \partial_{\nu} B_{\nu} \right) \right]^{(1)} \\
  {\cal F}_i^{(2)} = \int_0^{\infty} F_{\nu, i}^{(2)} \, d\nu \\
  = \left[ - \frac{a_{\text{\tiny R}} \, c}{3 \, \sigma_{\text{t}, \text{\tiny R}}} \, \partial_i T^4 + \frac{4}{3} \, u_i \, a_{\text{\tiny R}} \, T^4 \right]^{(1)} \, .
\end{multline}
% \end{linenomath}
\end{subequations}
This completes the calculations that are used in Section \ref{sec:asymptotic_analysis}.
Explicit expressions for ${\cal E}^{(2)}$ and ${\cal P}_{ij}^{(2)}$ cannot be produced because the functional form of the cross sections is not known.
The zeroth and second angular moments of $B_{\nu}^{(2)}$ produce the EDA-like results for ${\cal E}^{(2)}$ and ${\cal P}_{ij}^{(2)}$.
However, these angular moments of the zeroth-order contribution to $I_{\nu}^{(2)}$ produce additional terms beyond the EDA expressions, which we have called transport corrections.
Castor \cite{Castor2007} provides some discussion of what these terms might mean physically.
%% \section{}
%% \label{}

\section*{References}
\bibliographystyle{elsarticle-num}
\bibliography{References}
%% References
%%
%% Following citation commands can be used in the body text:
%% Usage of \cite is as follows:
%%   \cite{key}          ==>>  [#]
%%   \cite[chap. 2]{key} ==>>  [#, chap. 2]
%%   \citet{key}         ==>>  Author [#]

%% References with bibTeX database:

%% \bibliographystyle{model3-num-names}
% \textbf{References} \\
% \bibliographystyle{elsarticle-num}
% \bibliography{References}

%% Authors are advised to submit their bibtex database files. They are
%% requested to list a bibtex style file in the manuscript if they do
%% not want to use model3-num-names.bst.

%% References without bibTeX database:

% \begin{thebibliography}{00}

%% \bibitem must have the following form:
%%   \bibitem{key}...
%%

% \bibitem{}

% \end{thebibliography}
% \end{comment}
\end{document}